\documentclass[twocolumn]{aastex631}
\usepackage{CJKutf8}
\newcommand{\cntext}[1]{\begin{CJK*}{UTF8}{bsmi}#1\end{CJK*}}
\newcommand{\rej}{R}
\newcommand{\nonrej}{NR}

\begin{document}

\title{The Effect of Donor Star Rejuvenation on Common Envelope Evolution}

\correspondingauthor{C. Landri}
\email{camille.landri@utf.mff.cuni.cz}

\author[0000-0001-8078-0905]{C. Landri}
\affiliation{Institute of Theoretical Physics, Faculty of Mathematics and Physics, Charles University, \\
V Holešovickách 2, CZ-180 00 Praha 8, Czech Republic}

\author[0000-0002-5294-0630]{P. M. Ricker}
\affiliation{Department of Astronomy and Illinois Center for Advanced Studies of the Universe, \\
University of Illinois,
1002 W. Green St., Urbana IL 61801}

\author[0000-0002-6718-9472]{M. Renzo}
\affiliation{Steward Observatory, University of Arizona, 933 N. Cherry Ave., Tucson, AZ 85721, USA}

\author[0000-0003-4692-5941]{S. Rau (\cntext{饒孝節})}
\affiliation{Department of Astronomy, University of Illinois,
1002 W. Green St., Urbana IL 61801}

\author[0000-0003-1817-3586]{A. Vigna-G\'omez}
\affiliation{Max-Planck-Institut f\"ur Astrophysik, Karl-Schwarzschild-Str. 1, D-85748 Garching, Germany}

\begin{abstract}
In close binary star systems, common envelope evolution may occur after a previous phase of mass transfer. Some isolated formation channels for double neutron star binaries suggest that the donor of common envelope evolution was the accretor of a previous phase of stable mass transfer. Accretion should substantially alter the structure of the donor, particularly by steepening the density gradient at the core-envelope interface and rejuvenating the star. We study the common envelope evolution of a donor that was the accretor of a previous phase of stable mass transfer and has a rejuvenated structure. We perform 3D hydrodynamics simulations of the common envelope evolution of a 18 $M_\odot$ supergiant with a 1.4 $M_\odot$ companion using rejuvenated and non-rejuvenated 1D stellar models for the donor. We compare the two simulations to characterize the effect of the rejuvenation on the outcome of the common envelope phase and the shape of the ejecta. We find that accounting for a previous phase of mass transfer reduces the duration of the inspiral phase by a factor of two, likely due to the different structure in the outer layers of the donor. In the rejuvenated case, the simulations show more equatorially concentrated and asymmetric ejecta, though both cases display evidence for the formation of a pressure-supported thick circumbinary disk. During the dynamical inspiral phase, the impact of rejuvenation on the unbinding of the envelope is unclear; we find that rejuvenation decreases the amount of unbound mass by 20$\%$ to 40$\%$ depending on the energy criterion used.

\end{abstract}

\keywords{Common envelope binary stars (2156) --- Common envelope evolution (2154) --- Hydrodynamics (1963) --- Hydrodynamical simulations (767) --- Interacting binary stars (801)}

\section{Introduction} \label{sec:intro}
An important fraction of stars is found in binary systems \citep{abt_multiplicity_1976, bonnell_massive_2004, duchene_stellar_2013, moe_mind_2017,offner_origin_2023}, especially for massive stars \citep{mason_high_2009,sana_binary_2012}. These binary systems show a large variety of configurations, ranging from very wide systems with separations of thousands of AU to binaries separated by a couple of solar radii. In the case of close binaries, with a separation of $\lesssim1$~AU, the proximity of the stars generally gives rise to interactions via mass transfer, and their evolution will diverge from that of a single star \citep[e.g.,][]{podsiadlowski_presupernova_1992, sana_binary_2012, langer_presupernova_2012, smith_mass_2014, de_marco_dawes_2017}. In particular, close stellar binaries commonly go through phases of stable mass transfer via Roche lobe overflow (RLOF). During this phase, the donor star overfills its Roche lobe, and material leaves the potential well of the donor to be accreted by the companion, altering its structure in various ways \citep{packet_spin-up_1981, cantiello_binary_2007, renzo_evolution_2021}. A massive main sequence (MS) accretor is expected to experience a rejuvenation process during which its core expands and increases its hydrogen content through convective mixing \citep{hellings_phenomenological_1983}. Besides increasing the lifetime of the accretor, this process also alters the structure of its core-envelope boundary (CEB) region. Recently, \cite{renzo_rejuvenated_2023} found that the rejuvenation process lowers the binding energy of the CEB, which is of particular interest if the system later undergoes common envelope evolution (CEE) with the rejuvenated star as the donor.

CEE is a phase of binary evolution during which the secondary star plunges into the envelope of a giant primary and orbits its core \citep[e.g.,][]{paczynski_common_1976,ivanova_common_2013,ivanova_common_2020,ropke_simulations_2023}. The drag exerted on the companion by the envelope causes the orbit to decay as the companion transfers energy and angular momentum to the envelope, potentially unbinding it. The outcomes of CEE depend on the envelope ejection: a successfully ejected envelope should allow the orbit decay to slow until the system stabilizes as a short-period binary, which may become a source of transient phenomena such as cataclysmic variables \citep{paczynski_common_1976} or X-ray binaries \citep[e.g.,][]{Kalogera_formation_1998}, or a double-degenerate binary that may become progenitors of gravitational wave events \citep[e.g.,][]{klencki_it_2021,marchant_role_2021}. On the other hand, if part of the envelope remains bound, the two stars may eventually merge. The merger may be observed as a luminous red nova \citep[e.g.,][]{soker_violent_2006,ivanova_identification_2013,pejcha_cool_2016} or become a Thorne-Z\.{y}tkow object \citep{thorne_red_1975,thorne_stars_1977} if the companion is a neutron star. Thus, CEE is an important process of binary evolution. It is, however, still unclear which physical processes are relevant during the inspiral phase of CEE \citep[e.g.,][]{nandez_recombination_2015,ohlmann_hydrodynamic_2016,macleod_bound_2018,reichardt_impact_2020,sand_common-envelope_2020,lau_common_2022}.

Since the outcome of CEE strongly depends on the unbinding of the envelope, the fact that rejuvenation lowers the binding energy of the CEB is of particular interest. The CEB is the region where the binding energy is greatest; therefore, a CEE with a rejuvenated donor might eject a larger part of the envelope and significantly influence the post-CEE separation and stellar masses. This scenario is of interest for formation channels of binary neutron stars that expect a first phase of stable mass transfer in which the accretor will later become the donor of a CEE phase \citep[e.g.,][]{tutukov_merger_1993, belczynski_first_2016,tauris_formation_2017,nathaniel_population_2024}. In such cases, the rejuvenation of the future donor of the CEE could alter the outcome of the CEE phase and the subsequent evolution of the binary.

In this paper, we investigate how a previous phase of mass transfer and the ensuing rejuvenation of a CEE donor impact the inspiral phase of CEE. To do so, we perform 3D hydrodynamics simulations of the inspiral phase of the CEE with rejuvenated and non-rejuvenated donor models of similar masses obtained by \cite{renzo_rejuvenated_2023}. We briefly describe the stellar models for the donor star as well as the setup of our hydrodynamics simulations in \S\ref{sec:methods}. We present the results of our simulations in \S\ref{sec:results} and discuss their implications in \S\ref{sec:discussion} before concluding in \S\ref{sec:conclusions}.

\section{Numerical methods} \label{sec:methods}

\subsection{1D stellar models}
\label{sec:mesa}

The donor star models used in this study are 1D models made under the assumption spherical symmetry. The 1D profiles of gas quantities are taken from the models computed by \cite{renzo_rejuvenated_2023} using MESA (version 15140; \citealt{paxton_modules_2011, paxton_modules_2013, paxton_modules_2015, paxton_modules_2018, paxton_modules_2019}). We use their non-rotating 17.84~$M_\odot$ single star for the non-rejuvenated donor and their 15~$M_\odot$ accretor that reaches 17.41~$M_\odot$ after case-B mass transfer for the rejuvenated donor, which are both on the higher end of the mass distribution for double NS progenitors \citep{vigna-gomez_common_2020}. Both donors have a radius of 500~$R_\odot$ and a metallicity of 0.1~$Z_\odot$, and they still possess helium-rich cores.

\begin{figure}[h]
    \centering
    \includegraphics[width=1\columnwidth]{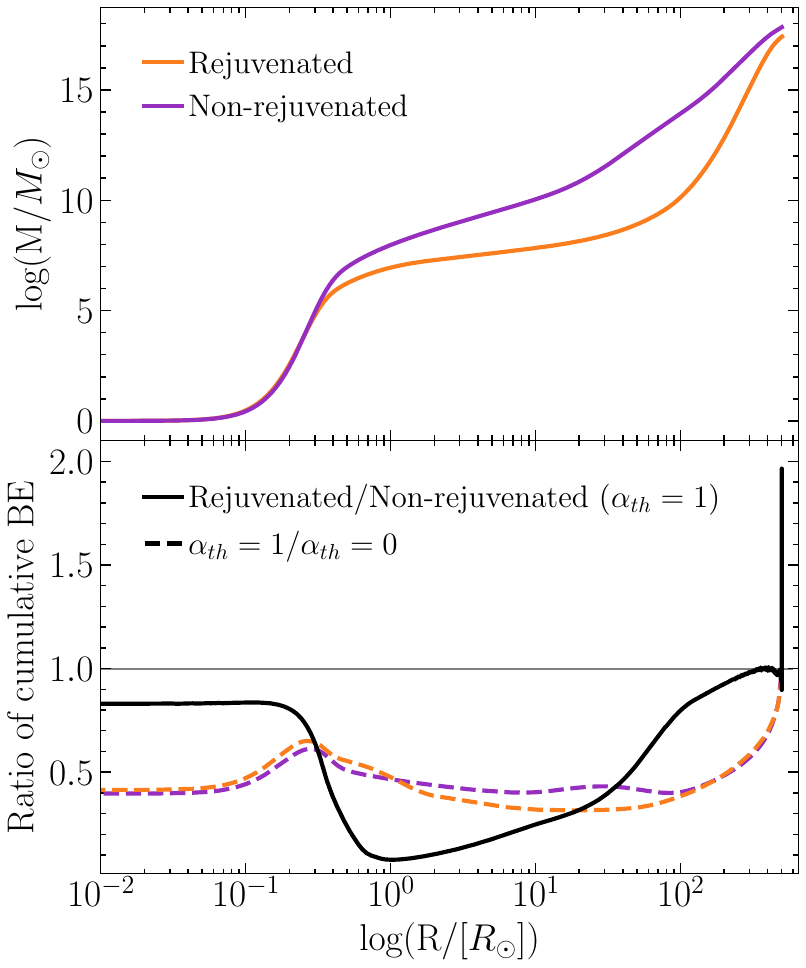}
    \caption{MESA stellar models for the rejuvenated (orange) and non-rejuvenated (purple) stars. Top panel: Enclosed mass. Bottom panel: Ratio of the cumulative BE. The black line shows the ratio of BE of the rejuvenated star to that of the non-rejuvenated star, using Equation~(\ref{eq:BE}) and assuming $\alpha_{\rm th}=1$. The dashed orange and purple lines show the ratio of BE with $\alpha_{\rm th}=1$ to BE with $\alpha_{\rm th}=0$ for the rejuvenated and non-rejuvenated models, respectively.}
    \label{fig:BE}
\end{figure}

In Figure~\ref{fig:BE} we show the differences in cumulative mass and binding energy (BE) of the two stellar models \citep[cf.\ Figures 2 and 3 in][]{renzo_rejuvenated_2023}. The binding energy was calculated for an enclosed mass $m$ and total mass $M$ using
\begin{equation}\label{eq:BE}
    {\rm BE}(m;\alpha_{\rm th}) = - \int^M_m \left(-\frac{Gm'}{r(m')}+\alpha_{\rm th}u(m')\right)dm'\ ,
\end{equation}
where $0\leq\alpha_{\rm th}\leq1$ is the fraction of internal energy that can be used to unbind the envelope, and is assumed to be equal to one here. The enclosed mass distribution is denoted by $m'(r)$, and the specific internal energy by $u(m')$. The value of the parameter $\alpha_{\rm th}$ impacts the outcome of the CE phase, since a larger contribution of the internal energy will help unbinding the envelope. In the bottom panel of Figure~\ref{fig:BE}, we show the difference in BE for each model when considering $\alpha_{\rm th}=0$ and $\alpha_{\rm th}=1$. By comparing these differences with the ratio of BE of rejuvenated to non-rejuvenated models, we see that rejuvenation is the main cause for the BE difference in the innermost layers, close to the CEB, and the value of $\alpha_{\rm th}$ mostly matters in the core and in the outer layers of the envelope. We thus expect the effect of rejuvenation to be dominant in the inner layers of the envelope, but it should be less significant compared to the impact of the choice of $\alpha_{\rm th}$ in the rest of the star.

\subsection{Initializing 3D hydrodynamical simulations}
\label{sec:flash_init}

\begin{figure}
    \centering
    \includegraphics[width=\columnwidth]{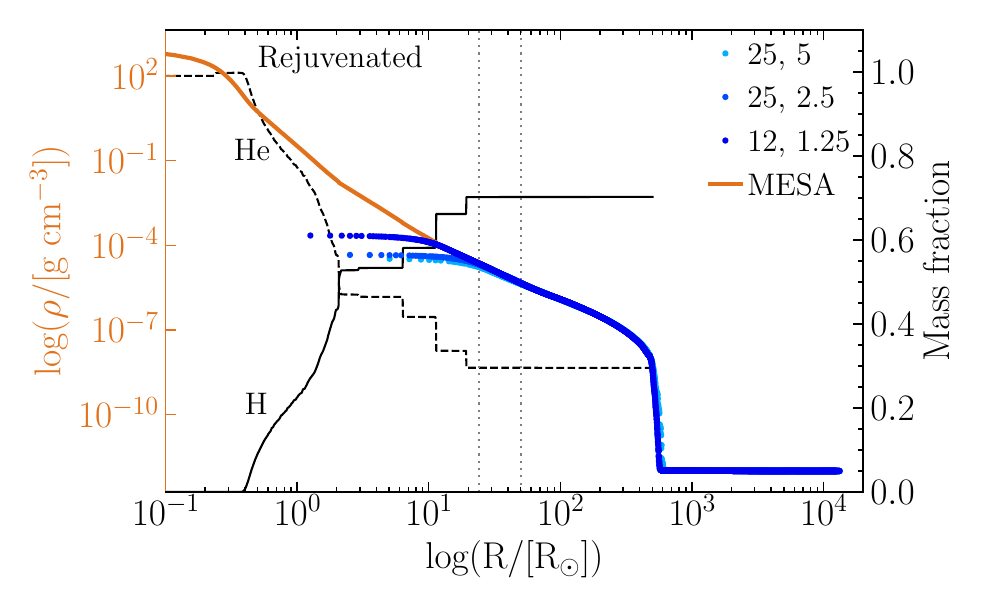}
    \includegraphics[width=\columnwidth]{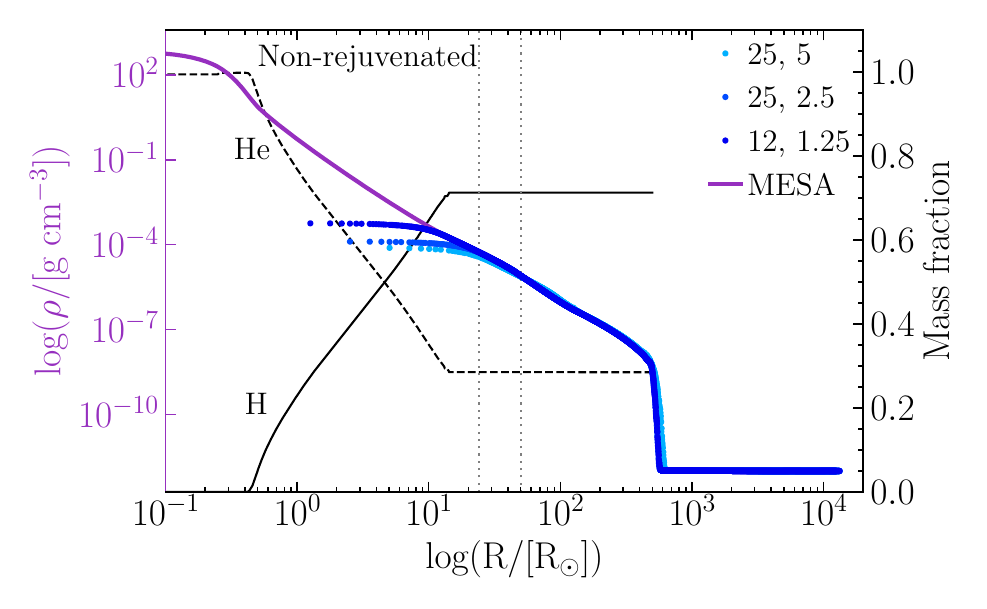}
    \caption{Density profiles as a function of radius of the donor at the start of the CEE. Runs are labeled using their numerical core radius and finest resolution. Upper panel: rejuvenated donor. Lower panel: non-rejuvenated donor. Blue dots represent the density profile for different resolutions and orange/purple lines represent the initial MESA profile. The black solid and dashed lines show the mass fractions of H and He respectively. The gray dotted line shows the sum of the core radii $2R_{\rm core}=50R_\odot$ and $2R_{\rm core}=24R_\odot$.}
    \label{fig:profiles}
\end{figure}

Our 3D common envelope simulations make use of FLASH 4.5 \citep{fryxell_flash_2000,dubey_introduction_2008}, an adaptive mesh refinement (AMR) hydrodynamics code. FLASH solves the Eulerian hydrodynamics equations on a mesh composed of a large number of Cartesian blocks, each of which contains the same number of zones. The blocks have different sizes and are organized in an oct-tree structure to increase resolution in regions of interest. The physics included and mesh refinement used are described further in \S\ref{sec:flash}.

To initialize each simulation, we interpolate one of the 1D MESA models described in \S\ref{sec:mesa} onto the AMR mesh and add a collisionless particle representing the companion star at the desired initial separation. To accurately capture the evolution of the CEE, we require the AMR mesh to cover a large enough volume to include all or most of the envelope as it expands away from the donor. This makes it difficult to adequately resolve the stellar core with available computing resources. For example, in our highest-resolution runs described below, the computational volume spans 72~AU on a side, while the smallest zone spacing is $1.26R_\odot$, so a uniform mesh would require $12,280^3$ zones and a correspondingly small timestep ($\sim 30$~minutes). Even with this resolution we just begin to resolve the core-envelope boundary. AMR makes these simulations tractable at this resolution, but fully resolving the stellar core would be prohibitively expensive.

To alleviate the extreme constraints on the resolution of the stellar core, we replace it with a collisionless numerical core particle following a procedure similar to \cite{ohlmann_constructing_2017}. The numerical core is assigned a radius $R_{\rm core}$ that can be resolved on the mesh, but it cannot represent all of the mass within $R_{\rm core}$, as that would leave an empty region at the center of the star without pressure support. We must therefore alter the MESA model within $R_{\rm core}$ and self-consistently determine the numerical core mass $M_{\rm core}$.

To do this, we join the MESA model outside $R_{\rm core}$ to a simplified hydrostatic model inside $R_{\rm core}$. This simplified model is a solution of the modified Lane-Emden equation representing a gaseous polytrope in the potential of a uniform-density spherical core of mass $M_{\rm core}$, which satisfies
\begin{equation}
    {{d}\over{d\xi}}\left(\xi^2{{d\theta}\over{d\xi}}\right) + \xi^2(\theta^n + \theta_c^n) = 0\ .
\end{equation}
Here we adopt the customary definitions for the density and radius variables $\theta$ and $\xi$ via
\begin{equation}
    \rho \equiv \rho_0 \theta^n\ ,\ \ \ r\equiv \alpha\xi\ ,
\end{equation}
where $\rho_0$ is the central density, $n$ is the polytropic index, and $\alpha$ is the scale height. This equation differs from the normal Lane-Emden equation \citep{lane_theoretical_1870,emden_gaskugeln_1907} by one term involving
\begin{equation}
    \theta_c^n \equiv {{M_{\rm core}}\over{4\pi R_{\rm core}^3 \rho_0}}\ .
\end{equation}
The density $\rho(R_{\rm core})$ and pressure $P(R_{\rm core})$ from the MESA model are used to determine the polytrope's specific entropy $K\equiv P/\rho^\gamma$ and thus also $\alpha^2 = [K(n+1)/(4\pi G)]\rho_0^{1/n-1}$. The polytropic index $n$ is set using the adiabatic index $\gamma$ reported by the equation of state at $r=R_{\rm core}$ via $n = 1/[\gamma(R_{\rm core})-1]$. Isotopic abundances are also matched to the MESA model at $R_{\rm core}$ and held constant throughout the polytrope. The core mass $M_{\rm core}$ and polytrope central density $\rho_0$ are varied in a nested pair of bisection loops until a solution that matches the MESA density and enclosed mass at $R_{\rm core}$ is found. The part of the MESA model inside $R_{\rm core}$ is replaced by the polytrope, and the derived core mass is used to initialize the numerical core particle at the center of the star.

Once the core properties have been determined, we linearly interpolate the density, pressure, and hydrogen, helium, and carbon/oxygen abundances from the modified MESA model onto the AMR grid in FLASH, centering the star within the domain and adding the numerical core particle. As described in \S\ref{sec:flash}, in the FLASH simulations we use the Helmholtz equation of state, which differs from the MESA equation of state in partially ionized regions. To preserve the hydrostatic equilibrium of the model, we interpolate the MESA pressure onto the AMR mesh and use the Helmholtz equation of state to set the temperature and energy of the gas. Outside the star, the gas is initialized as a uniform ``fluff'' medium at rest with temperature and density set to approximately balance any outflow due to unresolved pressure gradients at the stellar surface (25000~K and $10^{-12}$~g~cm$^{-3}$). To facilitate analysis, a passively advected tracer fluid is initialized to 1 in fluff regions and 0 inside the star. All velocities are initialized to zero at this stage.

After initialization, we allow the donor to relax for $10t_{\rm ff,donor}$, where $t_{\rm ff,donor} = [4\pi R_{\rm donor}^3/(3GM_{\rm donor})]^{1/2}$ is the free-fall timescale for a donor of mass $M_{\rm donor}$ and radius $R_{\rm donor}$. All the gas initialized on the grid, including the polytrope inside $R_{\rm core}$, is regarded as ``envelope'' material with total mass $M_{\rm env} = M_{\rm donor} - M_{\rm core}$. During this relaxation period we damp the velocity field by multiplying all velocities by 0.9 at the end of each timestep. We show density profiles of the resulting rejuvenated and non-rejuvenated donor models in Figure~\ref{fig:profiles}.

We then restart the simulation from a checkpoint file, turn off damping, and add the companion star. The companion star is  initialized as a particle with the same radius $R_{\rm core}$ as the donor core, but with a mass $1.4M_\odot$ representing a neutron star. We place it on the $x$-axis at a separation $a_{\rm init}$ for which the Roche lobe radius (computed using the \citealt{eggleton_approximations_1983} approximation) equals the MESA model's stellar radius $R_{\rm donor}$. These initial separations correspond to $843.9R_\odot$ (period $654.6$~d or 1.79~yr) for the rejuvenated donor and $838.5R_\odot$ (period $641.1$~d or 1.76~yr) for the non-rejuvenated donor. We also reset the velocity of any gas with a density above the fluff density and a fluff tracer fraction less than 0.9 so that this gas is in solid-body rotation about the donor core with an angular velocity equal to 95\% of the corotation frequency. At the surface of the donor this corresponds to 45\% and 35\% of the critical rotation speed for the rejuvenated and non-rejuvenated donors, respectively.

The common envelope simulation is then run from this initial condition until the orbit stabilizes or the core separation falls below $2R_{\rm core}$. The core separation at this time ($t_{\rm final}$) is denoted by $a_{\rm final}$. We note, however, that a subsequent slow inspiral phase not followed here can modify this ``final'' separation further, and this slower phase may be sensitive to the rejuvenation status of the donor.

With this setup, the donor is not in equilibrium with the binary potential at the beginning, but because all runs are initialized the same way we expect the differences between them to be mainly due to the differences in the donor model. To test this assumption, we used a smoothed particle hydrodynamics (SPH) code designed for relaxing binary models in the corotating frame to compare the energies of the two binary systems under two configurations. In the first configuration, we mapped the 1D stellar models onto the SPH code at the orbital separations $a_{\rm init}$ used in this paper, then measured the kinetic, internal, and potential energies, including orbital motion and spin in the kinetic energy. In the second configuration, we mapped the models at $4a_{\rm init}$, allowed them to relax for $t_{\rm ff,donor}$, forced the binaries together to $a_{\rm init}$ over an interval of $8t_{\rm ff,donor}$, then allowed them to relax for an additional $t_{\rm ff,donor}$ before measuring the energies. In each case we used the (25, 2.5) resolution described below together with approximately 50,000 SPH particles, and the same core replacement algorithm was used as in the FLASH simulations. We found that for both donor models, the kinetic energy was 6 -- 7\% smaller in the unrelaxed configurations, while the internal energy was 10\% larger and the potential energy was 9 -- 12\% larger in magnitude. Since the fractional changes in the energies between configurations are very similar for the two donors and at the 10\% level, we expect differences larger than this to be due to the different donor structures.

\subsection{3D hydrodynamical simulations}
\label{sec:flash}

Within FLASH we use the directionally split piecewise parabolic method solver together with the Helmholtz equation of state, parallel FFT-based multigrid Poisson solver \citep{ricker_direct_2008,daley_optimization_2012}, and a second-order leapfrog time integrator for particles. Simulations take place within a 3D Cartesian volume 72~AU on a side, with ``diode'' boundary conditions for hydrodynamics and isolated boundary conditions for the gravitational field. Diode boundary conditions correspond to outflow (zero-gradient) boundaries unless the velocity component normal to the boundary is inwardly directed, in which case this component is set to zero in the boundary zones. This allows material to escape the domain but prevents any inward velocity that develops at the boundary from creating artificial inflows.

Each AMR block contains $8^3$ zones, and the coarsest level of refinement contains $12^3$ blocks. Blocks are refined by applying the default FLASH second-derivative criterion to the density and pressure and by requiring refinement of blocks containing any zone whose center is within $4R_{\rm core}$ of a stellar core. We allow all blocks that contain a stellar core to refine to a higher maximum refinement level than those that do not. Convergence testing showed that, for the donor models considered in this paper, we required the numerical core radius to be at least 5 and preferably 10 times the smallest zone spacing $\Delta x_{\rm min}$ in order to relax the donor without producing unphysically strong convective instabilities (this is more stringent than we have found in simulations with low-mass donors). To avoid excessive refinement of the volume, we force derefinement of blocks whose maximum density is smaller than $10^{-10}{\rm\ g\ cm}^{-3}$ or which lie outside a distance of 18~AU from the center of the computational volume. We find when enforcing this constraint that very low-density material at the boundary of the refined region sometimes develops standing waves parallel to the refinement boundary; these do not appear to affect the overall solution.

The particles representing the numerical core of the donor and the companion interact only gravitationally with the gas. The interaction is determined by computing the acceleration due to the two cores in each mesh zone and storing it as an AMR variable that is added to the finite-differenced gas potential found with the Poisson solver. The acceleration of each core due to the gas is summed during this loop in such a way as to ensure momentum conservation. The donor core and companion also experience a mutual gravitational interaction. This method differs from the technique used in previous published FLASH common envelope simulations \citep[e.g.,][]{ricker_taam_2012} and yields greatly improved conservation properties, though because the gas gravitational acceleration is not conservatively differenced in the Euler equations, and the time centering of the gas and particle update steps is not the same, we do not conserve momentum to within round-off error. We monitor conservation in each of the runs and present results from the highest-resolution cases in the Appendix.

For each stellar model we performed runs at several different resolutions, varying the maximum level of refinement and the number of finest-level zones per numerical core radius. Table~\ref{tab:runs} summarizes the different simulations, and Figure~\ref{fig:profiles} shows the density profiles after the replacement of the core for different resolutions, as well as the initial MESA density and composition profiles. In the rest of this paper we use the notation ${\rm \rej}(R_{\rm core},\Delta x_{\rm min})$ to refer to runs with a rejuvenated donor and ${\rm \nonrej}(R_{\rm core},\Delta x_{\rm min})$ to refer to runs with a non-rejuvenated donor.

\begin{table}
\movetableright -0.5in
\caption{Simulations performed for this study. $R_{\rm core}$ is the radius of the numerical core, $\Delta x_{\rm min}$ is the size of the smallest cell, $M_{\rm core}$ is the mass of the numerical core, and $M_{\rm env}$ is the mass of the envelope.} 
\label{tab:runs}
\begin{center}
\begin{tabular}{cccccc}
\hline\hline
Donor & $R_{\rm core}$ & $\Delta x_{\rm min}$ & $M_{\rm core}$ & $M_{\rm env}$ \\ 
      & ($R_\odot$) & ($R_\odot$) & ($M_\odot$) & ($M_\odot$)  \\
\hline
Rejuvenated        & 25 & 5.0 & 7.91 & 9.50 \\ 
   (\rej)         & 25 & 2.5 & 7.91 & 9.50 \\
            & 12 & 1.26 & 7.68 & 9.64  \\
           \hline
Non-rejuvenated    & 25 & 5.0 & 10.26 & 7.58  \\
  (\nonrej)      & 25 &2.5 & 10.26 & 7.58  \\
            & 12 & 1.26 & 9.68 & 8.06  \\
\hline
\end{tabular}
\end{center}
\end{table}

\begin{figure*}[h]
    \includegraphics[width=0.49\linewidth]{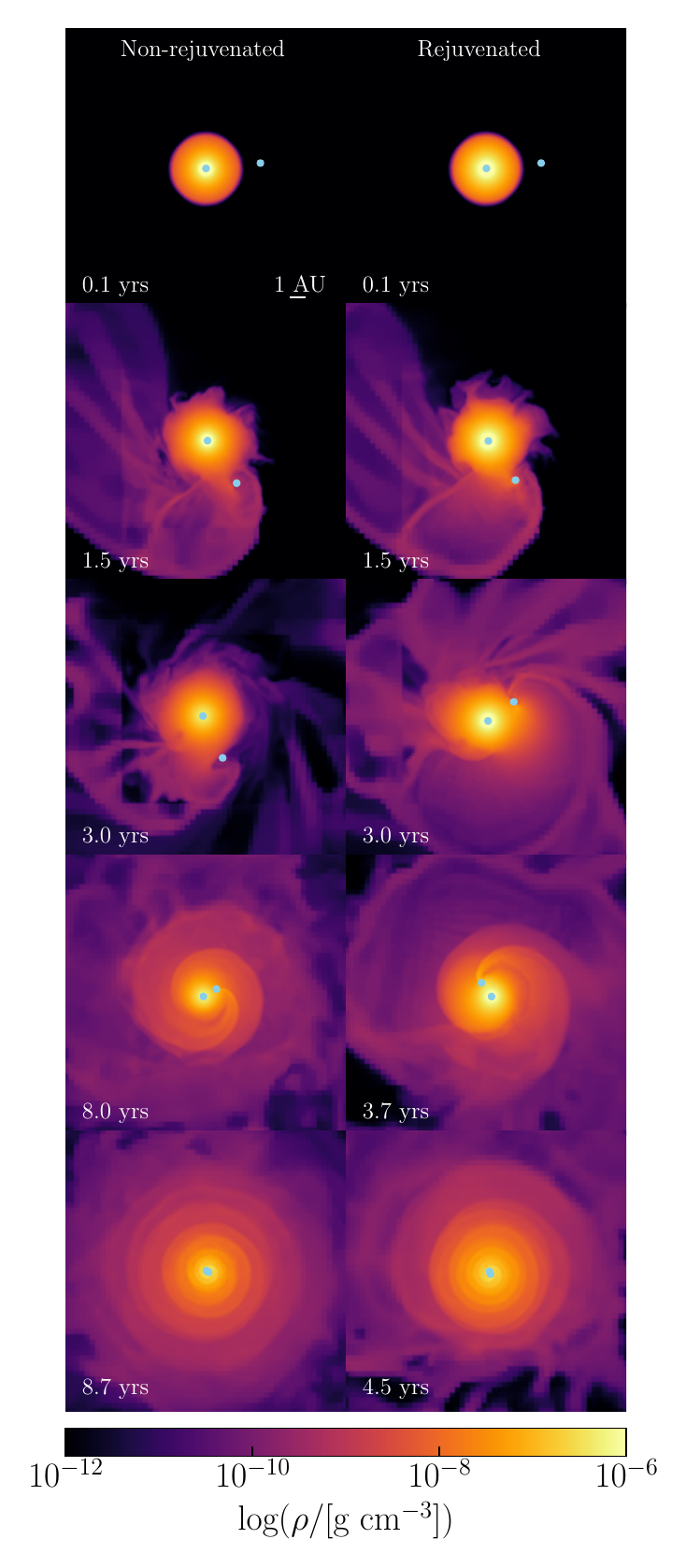} 
    \includegraphics[width=0.49\linewidth]{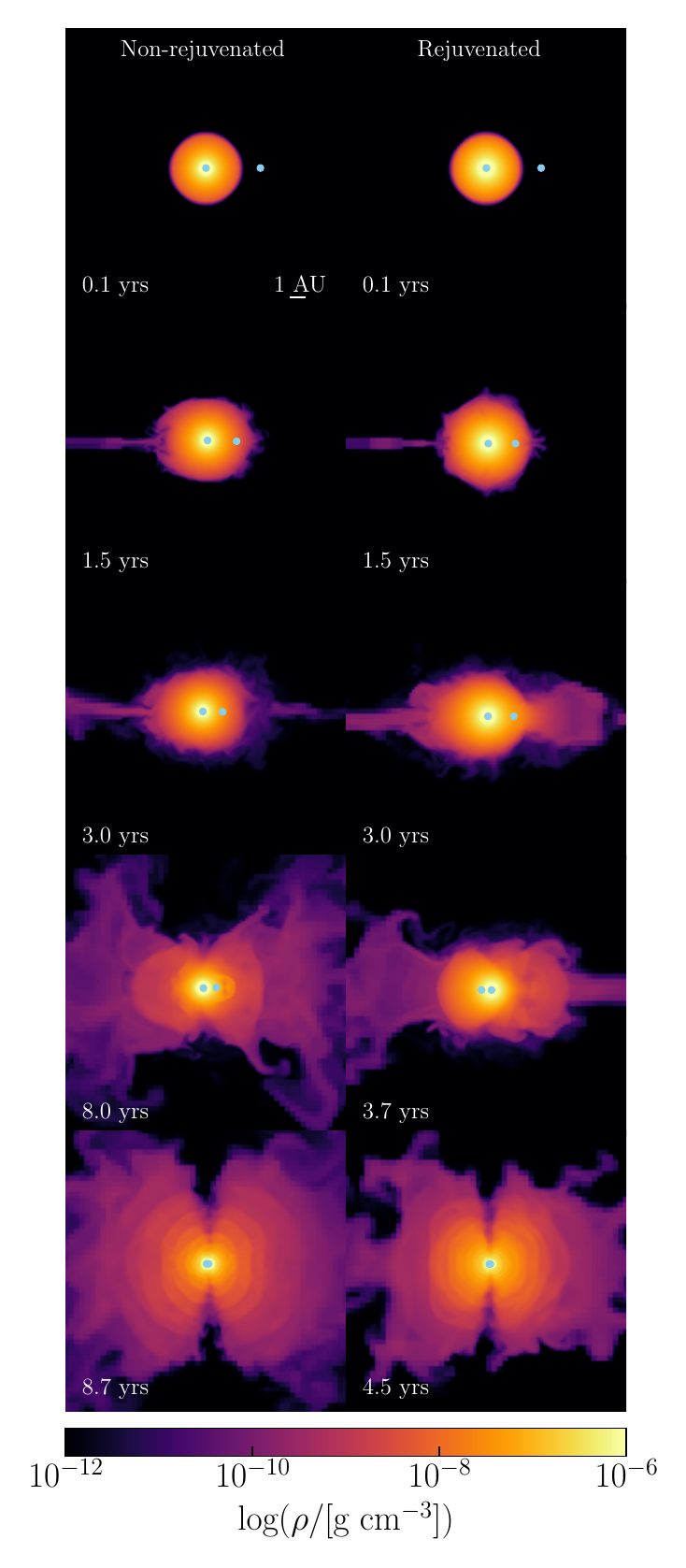}
    \caption{Snapshots of the \rej(12,~1.26) and \nonrej(12,~1.26) runs in the $xy$ (equatorial) plane (left) and in the $xz$ (meridional) plane (right), centered on the position of the donor's core. Each panel shows a density slice at $z=0$ and is 20~by~20~AU.  The donor core and companion are denoted as grey points. First row: start of the run. Second and third row: first grazing of the envelope of the donor by the companion. Fourth row: core-companion separation is around  1~AU. Fifth row: core-companion separation is $2R_{\rm core}$.}\label{fig:snap}
\end{figure*}

\section{Results}\label{sec:results}
We ran a total of six simulations, three for each donor model. We show an overview of the simulations in Figure~\ref{fig:snap} through density snapshots of the \rej(12,~1.26) and \nonrej(12,~1.26) models at different phases of the evolution of the system. In the early stages of each simulations (shown in rows 1 -- 2 of Figure~\ref{fig:snap}) the companion accretes mass from the donor while the orbit slowly shrinks. Then it grazes the outer envelope of the donor, until it plunges into the envelope and the dynamical inspiral starts (rows 3 -- 4). As the companion spirals further into the shared envelope, matter is lifted off the donor, and tight spiral patterns form at late stages (row 5). We summarize the final state of each run in Table~\ref{tab:results} and discuss the evolution of different quantities of interest throughout our simulations in the following subsections.

\begin{table}
\centering
\caption{Unbound mass and orbital separation at the final time in each simulation. } 
\label{tab:results}
\begin{tabular}{ccccc}
\hline\hline
Simulation & $t_{\rm final}$ & $M_{\rm K+P}$  & $M_{\rm K+P+Th}$  & $a_{\rm final}$ \\ 
      & (yr) & ($M_\odot$) & ($M_\odot$) &  ($R_\odot$) \\
\hline
\rej(25,5.0)  & 3.7 & 0.09 & 0.55 & $\leq50$ \\ 
\rej(25,2.5)  & 4.5 & 0.12  & 0.60 & $\leq50$ \\ 
\rej(12,1.26) & 4.3 & 0.13  & 0.57 & $\leq24$ \\ 
\hline
\nonrej(25,5.0)  & 5.7 & 0.08 & 0.66 & $\leq50$ \\
\nonrej(25,2.5)  & 8.9 & 0.09 & 0.75 & $\leq50$ \\
\nonrej(12,1.26) & 8.6 & 0.08 & 0.74 & $\leq24$ \\
\hline
\end{tabular}
\end{table}

\subsection{Evolution of the orbit} \label{sec:binary}

\begin{figure}[h]
    \centering
    \includegraphics[width=\columnwidth]{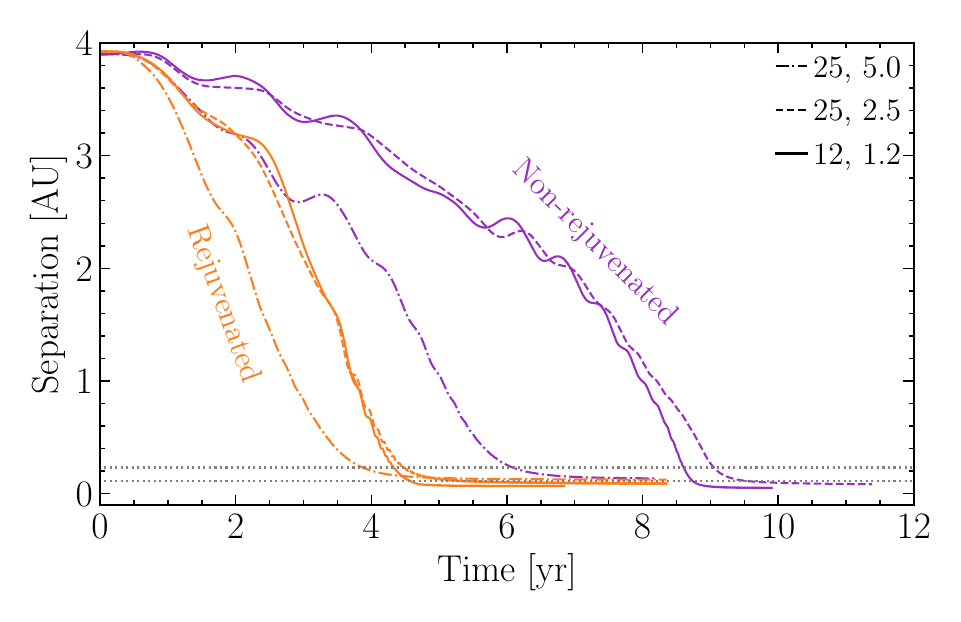}
    \caption{Evolution of the separation of the cores during the CEE for both rejuvenated (orange) and non-rejuvenated (purple) donors at different resolutions. The grey dotted lines denote where the separation is equal to the sum of the core radii, $2R_{\rm core}=50R_\odot$ and $2R_{\rm core}=24R_\odot$.}
    \label{fig:sep}
\end{figure}

We first examine the orbit of the binary during CEE by studying the evolution of the separation of the numerical cores, which is shown in Figure~\ref{fig:sep} for both rejuvenated and non-rejuvenated models. This evolution can be broadly divided into three phases. First, the orbit slowly loses stability and becomes eccentric as the companion accretes gas from the donor and spirals inward toward its surface. In the second phase, the companion grazes the outer envelope of the donor, and the orbital inspiral begins to accelerate. This phase appears to last longer for the non-rejuvenated donor. In the third phase the companion plunges into the envelope of the donor. The modulation of separation due to the eccentricity becomes smaller as the orbit decays, indicating an eventual circularization. Nevertheless, the binary remains eccentric during the whole inspiral phase, even after the orbit regains its stability near the resolution limit. The lower-resolution runs show that $R_{\rm core}/\Delta x_{\rm min} \gtrsim 10$ is needed to obtain converged results.

While we observe the same overall evolution for the two donor models, we note a change of pace in the inspiral between the two sets of simulations. The first and second phases of the orbit last around two years for rejuvenated donors, but it takes six years for the companion to start the fast inspiral in the envelope of the non-rejuvenated donor. This can be seen from the third row of Figure~\ref{fig:snap}, where the companion is already inside the envelope of the rejuvenated donor and drives significant outflows, while the non-rejuvenated system is still in the earliest phase, with less gas ejected. The speed of the fast phase of inspiral is also slightly affected: it takes about two years for the companion to reach a stable orbit in the rejuvenated donor, against almost three years in the non-rejuvenated case. As a result, the binary with a rejuvenated donor reaches the minimum separation between cores after 4.5~years, while the non-rejuvenated inspiral takes almost 9~years. We hypothesize that this faster pace is due to the difference in the structure of the outer layers of the rejuvenated model, as shown by the larger mass content and higher binding energy of the outer envelope (see Figure~\ref{fig:BE}).

Finally, we remark that while our simulations show stabilized orbits, the separation of the binaries has reached $2R_{\rm core}$ before the orbit has become stable. Considering that the density of the gas inside the domain delimited by the numerical cores is much lower than the density in the stellar profiles, the gravitational drag and torques between the companion and the gas are very likely underestimated. As such, we do not take any results after the cores have coalesced as physically meaningful and consider that the orbit of the cores did not manage to stabilize in any of our simulations. It is thus impossible to conclude whether the stellar cores will merge or stabilize in a tight orbit by the end of the inspiral phase, and we can only derive an upper limit on the final separation of the binary and an upper limit on the duration of the CE phase.

\subsection{Outflows}
\label{sec:outflows}
\begin{figure}
    \centering
    \includegraphics[width=\columnwidth]{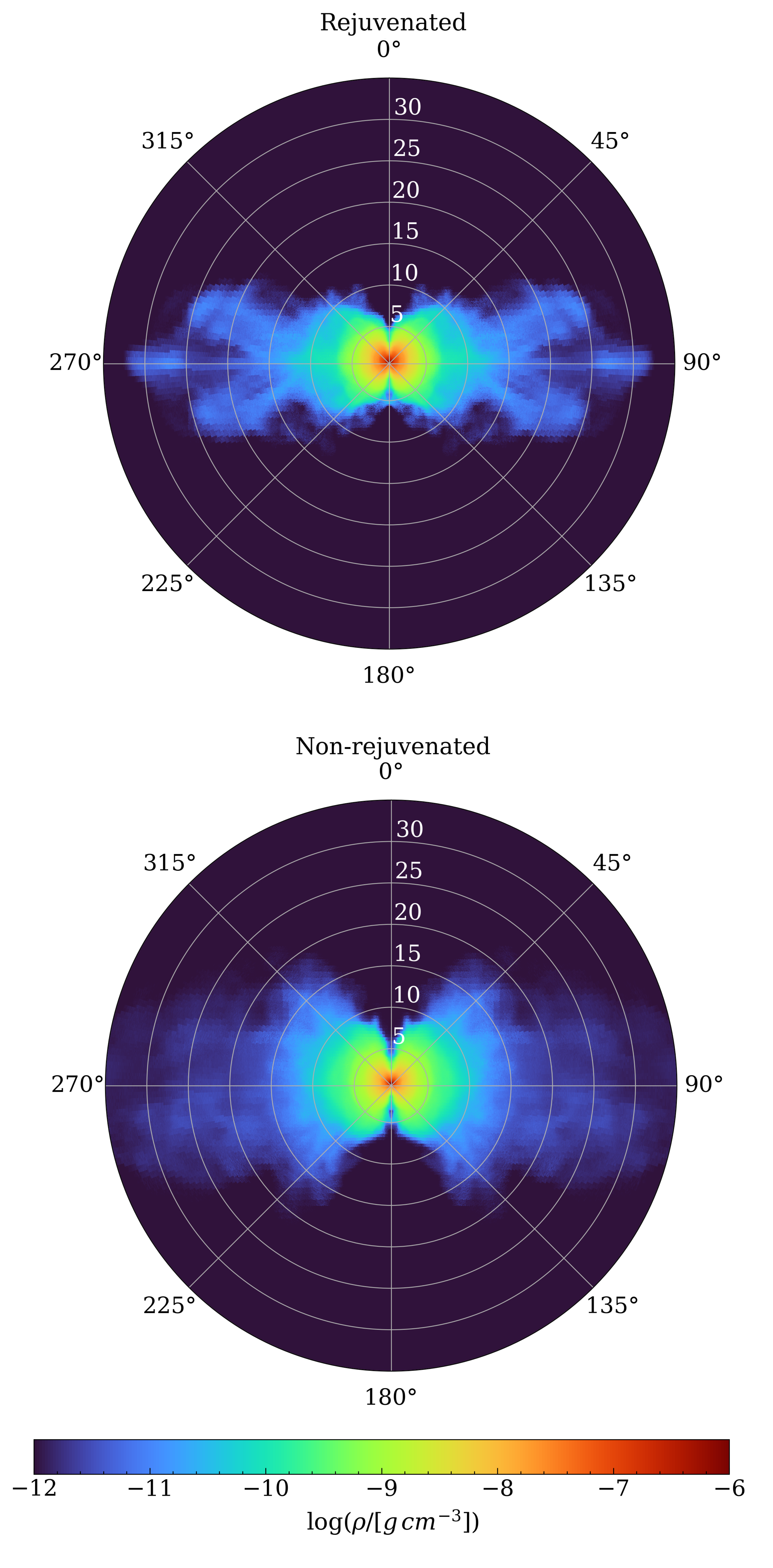}
    \caption{Azimuthally averaged density of the R(12,1.26) and NR(12,1.26) models in the top and bottom panels, respectively. Circles are labeled with radii in AU.}
    \label{fig:az_av}
\end{figure}

The evolution of the shape of the envelope during the high-resolution simulations is shown in Figure~\ref{fig:snap}. In the final snapshots, taken just before the cores coalesce, we see the expected over-densities due to the inspiral of the companion in both donor models, but the spiral pattern appears slightly tighter and more concentrated toward the center of the binary in the case of the rejuvenated donor. From the slices in both the equatorial and the meridional planes, we see that the non-rejuvenated envelope has spread further by the end of the simulation than the rejuvenated envelope. Additionally, the slices of the meridional plane of the envelope show strong spherical asymmetries, with a large decrease in density around the polar axis of both models. In particular, the matter distribution in the rejuvenated envelope appears significantly asymmetric with respect to the $z$-axis of the grid.

\begin{figure}
    \centering
    \includegraphics[width=\columnwidth]{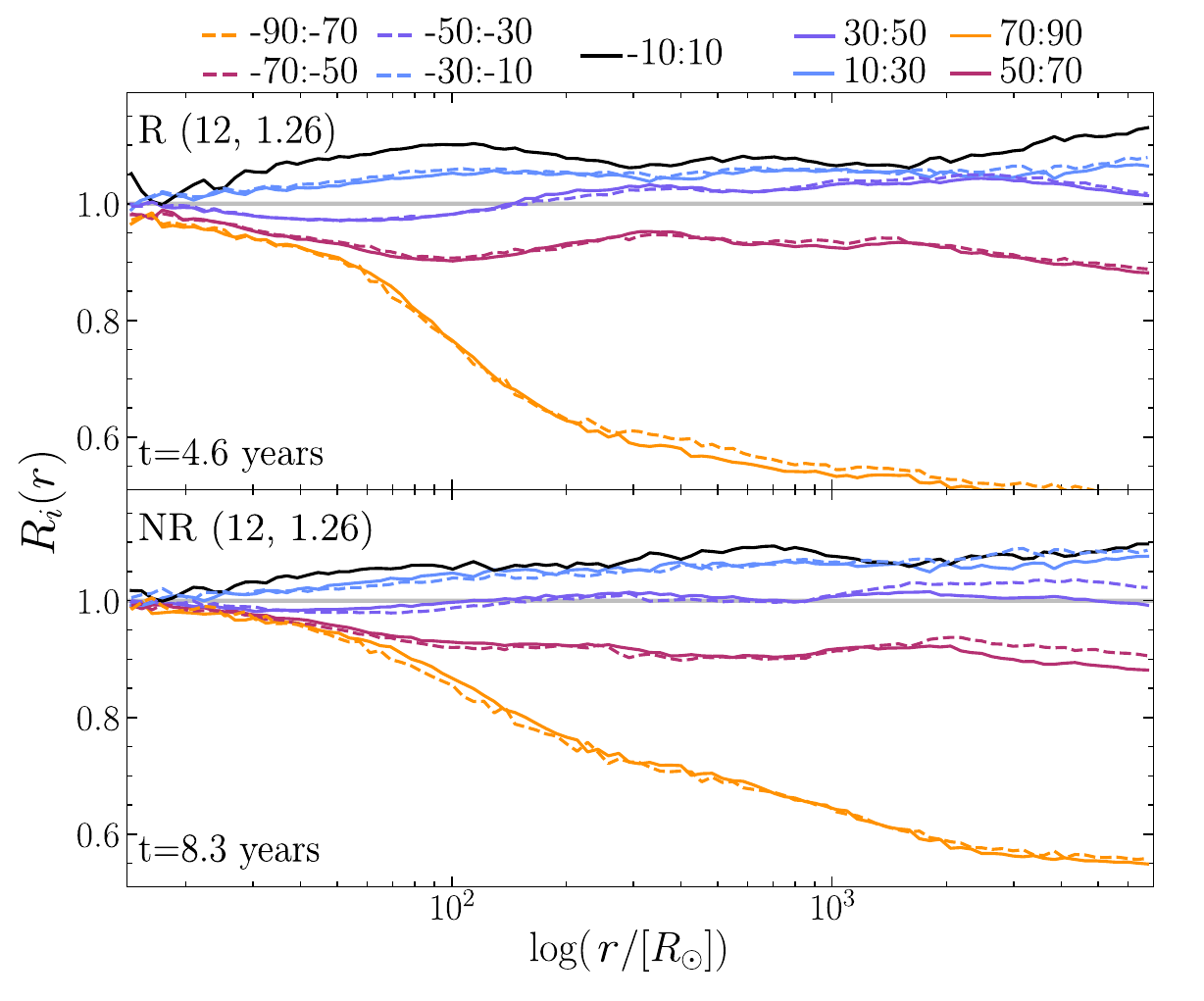}
    \caption{Relative mass distribution of the envelope within different radii of the donor core, as defined by Equation~(\ref{eqn:r}), at $t_{\rm final}$, for R(12,1.26) and NR(12,1.26) in the top and bottom panels, respectively. Each curve shows a different polar angle bin, with angle measured from the orbital plane. }
    \label{fig:mpolar}
\end{figure}

We inspect the shape of the ejecta and the envelope at the end of the simulation in Figures~\ref{fig:az_av} and \ref{fig:mpolar}. In particular, we are interested in the isotropy of the outflow. Figure~\ref{fig:az_av} shows the azimuthally averaged density (centered on the donor core) of the gas right before the cores coalesce in the case of the rejuvenated and non-rejuvenated (12,~1.26) models. Both models have non-spherical outflows with substantial under-densities in the polar regions, but the ejecta are more equatorially concentrated in the case of the rejuvenated donor. To quantify how isotropic the mass distribution is at different radii, we examine the ratio 
\begin{equation}
    R_i(r) \equiv {{\int_{\theta_i}^{\theta_{i+1}} \int_0^r \rho r^2\sin\theta\,dr\,d\theta}\over{\int_0^\pi \int_0^r \rho r^2\sin\theta\,dr\,d\theta}} {{\int_0^\pi \sin\theta\,d\theta}\over{\int_{\theta_i}^{\theta_{i+1}} \sin\theta\,d\theta}}
    \label{eqn:r}
\end{equation}
of enclosed mass within radius $r$ and polar angle range $[\theta_i, \theta_{i+1}]$ to that within radius $r$, divided by the corresponding ratio of volumes. This definition accounts for the fact that different polar angle bins have different enclosed volumes; if $\rho$ is a function of $r$ only (i.e.\ the mass distribution is isotropic but not necessarily uniform), then $R_i(r) = 1$ for all $i$ and $r$.
The result of this analysis, shown in Figure~\ref{fig:mpolar}, indicates that most of the ejected gas is concentrated in the range of polar angle $[-50^\circ,50^\circ]$ relative to the orbital plane, leaving the polar regions mostly empty beyond a few $\sim 100R_\odot$. In the case of the rejuvenated envelope, the asphericity of the gas is strong even in the inner regions of the ejecta ($R<50R_\odot$), whereas the mass distribution of the inner non-rejuvenated envelope is closer to isotropic.

\begin{figure*}
    \centering
    \includegraphics[width=0.95\linewidth]{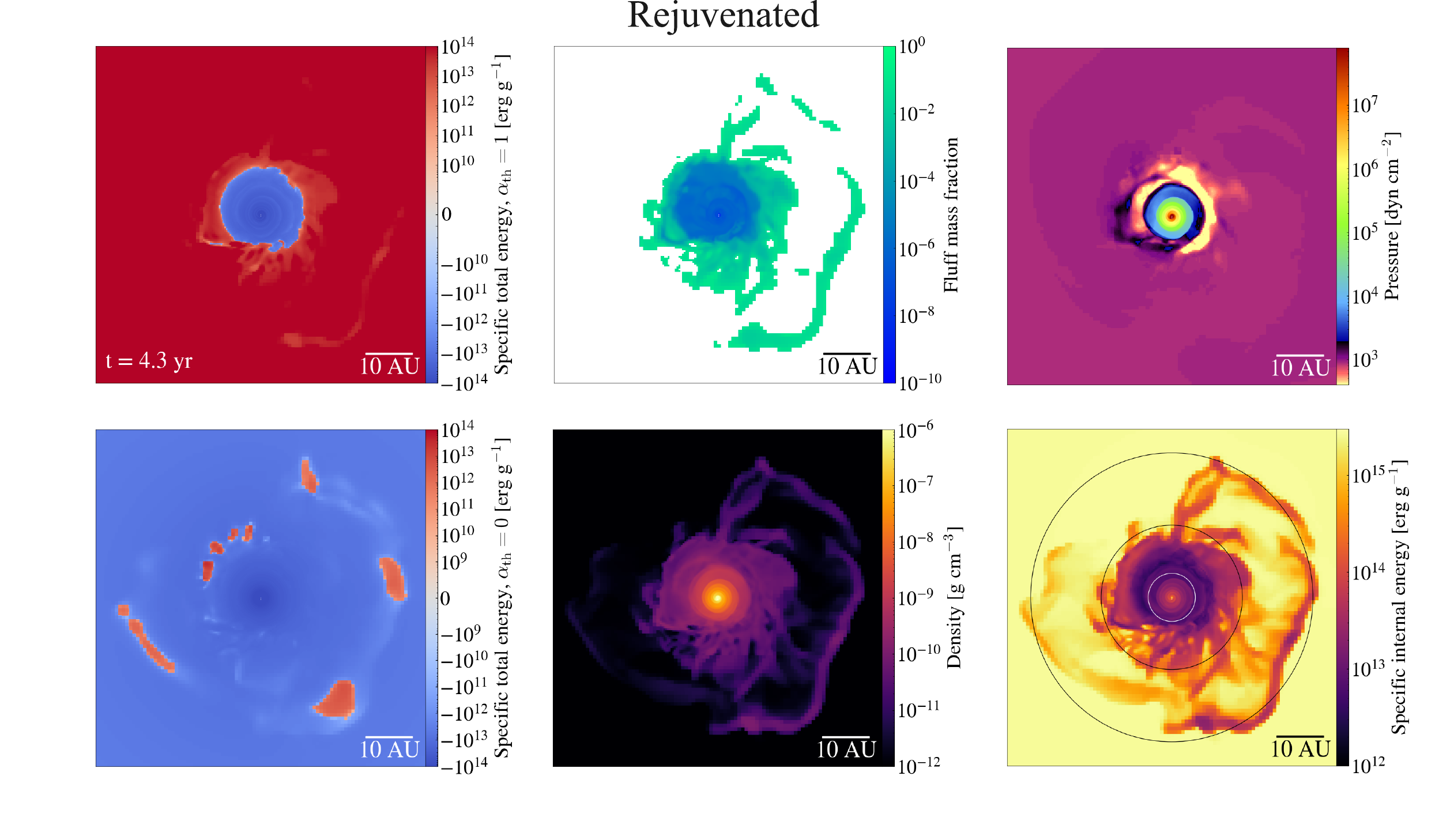}
    \includegraphics[width=0.95\linewidth]{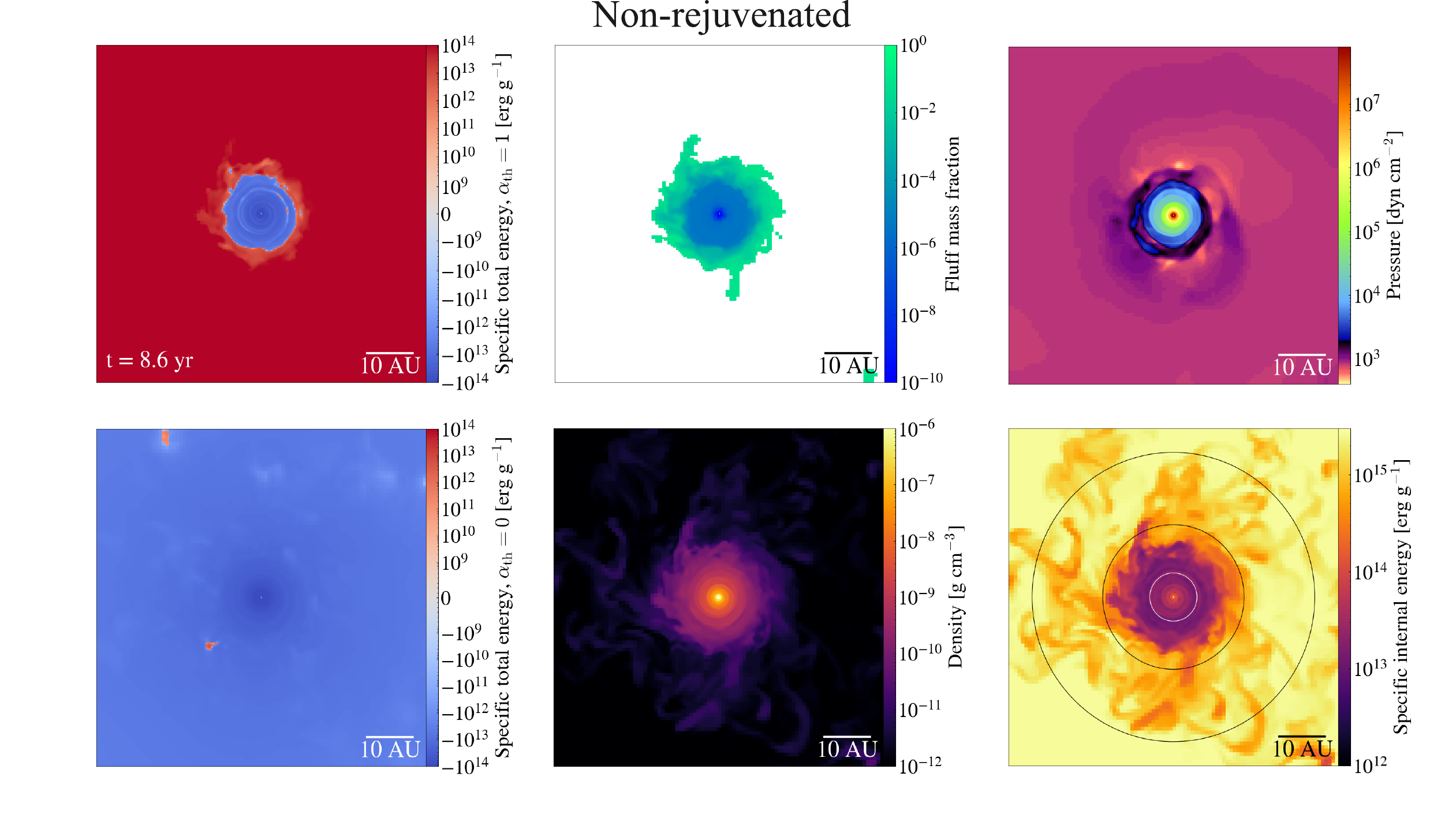}
    \caption{Orbital-plane properties of the \rej(12,1.26) run at $t = 4.3$~yr and the \nonrej(12,1.26) run at $t = 8.6$~yr, corresponding to when the separation drops below $R_\textrm{core}$. Left panels: specific energy with (top) and without (bottom) including internal energy. Red shades represent unbound material, while blue shades represent bound material. Center panels: mass fraction of the low-density fluff material for regions of fluff abundance below 0.1 (top) and density (bottom). Right panels: pressure (top) and specific internal energy (bottom). Black and white circles indicate boundaries of the three regions discussed in the text and Table~\ref{tab:support}.}
    \label{fig:sixpanel}
\end{figure*}

To gain further insight into the behavior of the envelope, we analyzed the role of pressure gradients and centrifugal acceleration in the final snapshots of the highest-resolution runs. Figure~\ref{fig:sixpanel} shows slices in the orbital plane (passing through the center of mass) of bound and unbound material, fluff mass fraction, and the gas density, pressure, and specific internal energy. We identify three distinct cylindrical regions in the specific internal energy slice: an inner region extending from 0 to 5~AU characterized by a tight spiral shock; a middle region extending from 5 to 15~AU characterized by smoother shocked flow; and an outer region extending from 15 to 30~AU characterized by low densities and shear instabilities and terminating in the irregular surface of the envelope. (We use the same radii for both donors although the outer radius of the middle region for the non-rejuvenated donor appears slightly smaller.) Mass-weighted spherical radius components of the gravitational acceleration ${\overline g}$, pressure acceleration $-{\overline {(\nabla p)/\rho}}$, and centrifugal acceleration ${\overline {v_\phi^2/r}}$ (based on cylindrical radius) for the three regions are listed in Table~\ref{tab:support}.

\begin{table}
\movetableright -0.6in
\caption{Mass-weighted averages of radial acceleration components at $t = t_{\rm final}$.} 
\label{tab:support}
\begin{tabular}{cccccc}
\hline\hline
Model & Region  & Mass & ${\overline g}$  & $-{\overline{(\nabla p)/\rho}}$ & ${\overline{v_\phi^2/r}}$ \\ 
 & & ($M_\odot$) &(cm\,s$^{-2}$) & (cm\,s$^{-2}$) & (cm\,s$^{-2}$) \\
\hline
\rej(12,1.26) & inner & 8.83 & $-9.5$ & 29 & 0.64 \\
& middle & 0.67 & $-0.20$  & 0.13 & 0.034 \\
& outer & 0.13 & $-0.028$ & 0.022 & 0.0061 \\
\hline
\nonrej(12,1.26) & inner & 7.09 & $-20$ & 88 & 1.30 \\
& middle & 0.86 & $-0.20$ & 0.19 & 0.032 \\
& outer &  0.014 & $-0.030$ & 0.040 & 0.0053 \\
\hline
\end{tabular}
\end{table}

The material in the outermost region originates during the early interval when the companion is outside the donor's envelope, and the orbital decay is driven by tidal interaction and the loss of material through the outer Lagrange points. During this phase the surface layers of the donor are transferred to the companion and expand within its Roche lobe, forming a contact discontinuity where they meet the hotter fluff material. This region shows some mixing with the fluff due to shear instabilities and a largely pressure-driven outflow (though the pressure gradient in this region is very small).

The middle region is produced during the part of the companion's first orbit in which it grazes the surface of the donor's envelope. The transferred gas accompanying the companion first interacts with the companion's wake during this phase, producing the first of the spiral shocks. While pressure support still dominates centrifugal support, the two balance the gravitational field in this region. Given the flattened ejecta structure, we could characterize this material as forming a circumbinary disk. The long-term interaction of this material with the stellar cores is likely important for determining the final separation \citep[similar arguments have been made for low-mass and very high-mass common envelopes;][]{kashi_circumbinary_2011,metzger_luminous_2022}, but we cannot follow this interaction with the current simulations.

In the inner region, a tight spiral shock pattern is produced as the companion spirals into the donor's envelope. This region shows a strongly pressure-driven outflow, feeding the middle disk region. If this outflow were to strengthen during further inspiral in a higher-resolution simulation, we might expect it to shock and unbind more of the material in the middle circumbinary disk region. However, in these runs it has not done so, contributing to the low rates of unbinding discussed in \S\ref{sec:massloss}.

\subsection{Mass loss}

\begin{figure}
    \centering
    \includegraphics[width=1\columnwidth]{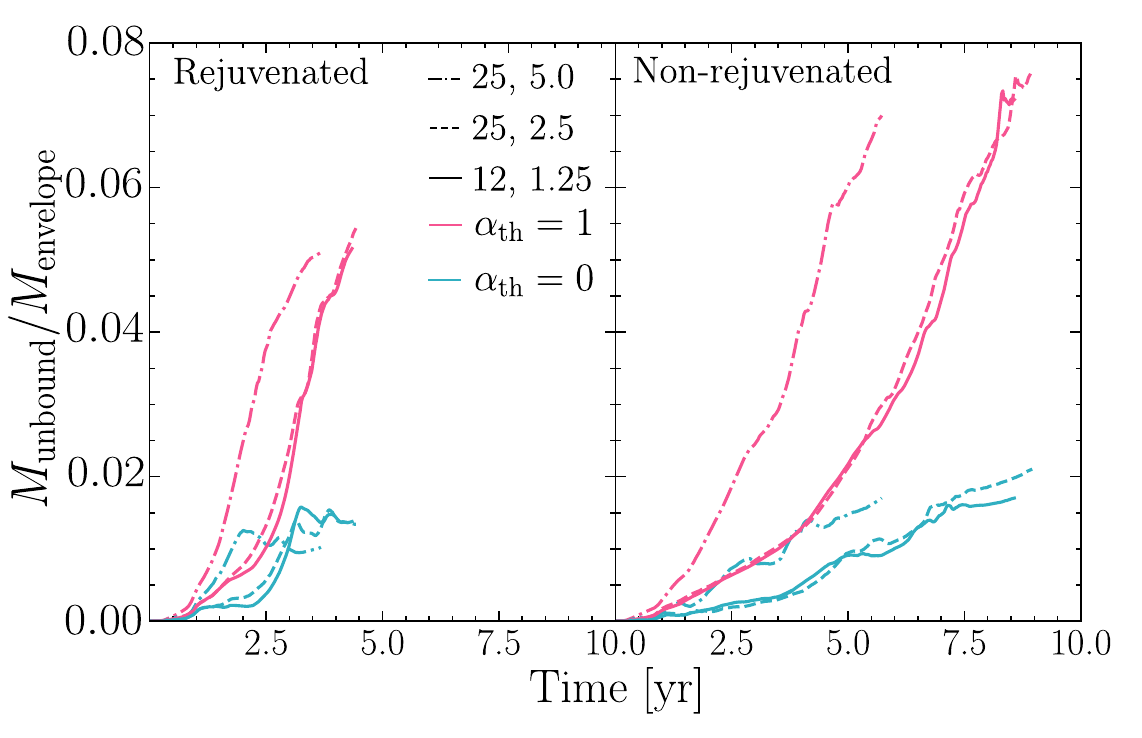}
    \caption{Evolution of the amount of unbound mass during the CEE for different resolutions and donor models. Cyan lines show the unbound mass considering the total energy is calculated using only the potential and kinetic energy of each cell ($\alpha_{\rm th}=0$). Pink lines show the unbound mass considering the total energy is calculated using the potential, kinetic and internal energy of each cell  ($\alpha_{\rm th}=1$). Mass is considered unbound if its total energy is positive or if it has left the domain. Left panel: Simulations with a rejuvenated donor. Right panel: Simulations with a non-rejuvenated donor.}
    \label{fig:mloss}
\end{figure}

\label{sec:massloss}
We show the fraction of envelope mass lost in our simulations as a function of time until the numerical cores coalesce in Figure~\ref{fig:mloss}. A mesh zone is considered unbound if its kinetic energy is greater than its potential energy:
\begin{equation}\label{eq:bound}
\phi + \frac{1}{2}v^2 > 0\ .
\end{equation}
Alternatively, we can assume that a fraction $\alpha_{\rm th}$ of the internal energy of the zone is converted into kinetic energy and consider the parcel unbound if
\begin{equation}\label{eq:boundth}
    \phi + \frac{1}{2}v^2 + \alpha_{\rm th} u > 0\ .
\end{equation}
We count mass lost from the computational domain as unbound, though in our simulations this amount is very small ($< 0.1\%$). 
The value of $\alpha_{\rm th}$, the parameter denoting what fraction of internal energy is used to unbind gas, is not well-determined by our simulations since they do not include radiative cooling or recombination energy and do not follow the long-term evolution of the common envelope after the dynamical phase. Setting it to unity provides an upper limit on the unbound mass. The more conservative criterion of Equation~(\ref{eq:bound}) serves as a lower limit. In both cases, the amount of mass lost is calculated using an upper limit on fluff abundance, such that any zone with a fluff abundance greater than 0.1 is not included in the sum. This cut removes a considerable fraction of the domain, including regions where the fluff has mixed with very small amounts of gas from the envelope. To get a better idea of which parts of the domain are taken into account for the mass loss, we use the middle panels of Figure~\ref{fig:sixpanel} to compare snapshots of the density and the fluff abundance of the gas with a 0.1 limit applied, so that white regions correspond to regions that have been cut out. We note that since all the binaries reach their minimum resolvable separation of $2R_{\rm core}$ in our simulations, a part of the inspiral is not captured. Higher resolution simulations with smaller $R_{\rm core}$ values should follow the inspiral deeper in the common envelope, possibly leading to more mass loss.

Figure~\ref{fig:mloss} shows that the rejuvenation of the donor does affect the amount of mass lost. When considering the less conservative criterion for mass loss (Equation~\ref{eq:boundth}), the rejuvenated giant loses up to 0.052$M_{\rm env}$ (0.50$M_\odot$) while the non-rejuvenated one loses up to 0.072$M_{\rm env}$ (0.58$M_\odot$). Neither donor model yields significant mass loss, though the fractional mass loss decreases by about 40\% when considering the rejuvenation of the donor. Similarly, when considering the more conservative criterion (Equation~\ref{eq:bound}), the mass loss in the rejuvenated system reaches 0.014$M_{\rm env}$ (0.13$M_\odot$) against 0.017$M_{\rm env}$ (0.14$M_\odot$) in the non-rejuvenated case, effectively showing a 20\% decrease when taking rejuvenation into account. In each case the low-resolution runs appear to converge to the high-resolution results for the simulation times covered. Unbinding is still occurring rapidly at the end of the runs when internal energy is included, and more slowly when it is neglected.

To investigate the origin of the difference in mass loss according to the two criteria, we show slices of bound and unbound material in the orbital plane according to the two criteria derived above in the leftmost panels of Figure~\ref{fig:sixpanel}. We see that adding internal energy to the binding criterion primarily affects the low-density outer region and the outer part of the middle region characterized as a circumbinary disk in \S\ref{sec:outflows}, causing them to become unbound. The inner region and inner part of the disk remain bound; thus, the difference between the two energy criteria is due to the significant internal energy of the outer regions of the ejecta. Given the dynamics of the inner region (expanding spiral shocks) and the additional orbital energy that might be released in higher-resolution simulations, we expect additional material in the disk to become formally unbound by shock heating in such simulations.

We cannot firmly conclude that one donor would lose more material than the other at the end of a full common envelope evolution. However, since some internal energy will be used to do work on the envelope, it appears that the rejuvenated donor will lose its envelope more rapidly than the non-rejuvenated donor, but that the total amount of envelope material lost may be considerably smaller. Both high-resolution simulations are consistent with either a merger of the cores or stabilization of the binary orbit below our resolution limit.

\section{Discussion} \label{sec:discussion}
Most 3D hydrodynamics simulations of CEE to this date have been performed for low-mass stars, though recently some effort has been put into massive CEE with NS companions. 

\cite{lawsmith.2020} performed simulations of a $12\,M_\odot$ donor and a $1.4\,M_\odot$ NS companion, and find a clean ejection of the envelope with a stabilized orbit at $\lesssim 5.1\,R_\odot$. However their methods strongly differ from ours, as they excise the outer layers of the donor and reduce it to a $10\,R_\odot$ star but resolve the core down to $\lesssim 0.005\,R_\odot$. In our case, the inspiral in the outer layers drastically changes between rejuvenated and non-rejuvenated models, suggesting that such a method might not be able to fully capture the effect of rejuvenation on CEE and thus would poorly reproduce the inspiral. \cite{moreno_3d_2022} performed magnetohydrodynamics simulations of a $10\,M_\odot$ with a $1.4\,M_\odot$ NS and found a successful envelope ejection and an orbit stalling at $15\,R_\odot$, indicating that our numerical cores are indeed too large to resolve the stabilization of the orbit, if it occurs. 

On the other hand, \cite{lau_common_2022} presented simulations of two cases, a $12\,M_\odot$ donor with a $1.4\,M_\odot$ companion (consistent with a NS) and a $3.0\,M_\odot$ companion (consistent with a black hole) to probe the effect of recombination energy. They found that the inspiral stalls only in the case of a black hole companion. Their simulations with the NS have a low mass ratio, similar to ours, and the orbit stalls after the numerical cores overlap, thus providing only an upper limit on the final separation, as in our case. In their simulations with black hole companions, they find that most of the envelope is unbound after the inspiral stalls, and the mass loss due to the dynamical inspiral is slightly larger than our results. Thus, if the binary stalls at a lower separation than we can currently resolve in our simulations, we may expect a significant increase in mass loss.

\subsection{Orbital decay and unbinding efficiency}
The main goal of this work was to investigate whether the significant difference in BE across the envelope of rejuvenated and non-rejuvenated donors, as shown in Figure~\ref{fig:BE}, would affect the inspiral phase of CEE and change the outcome of the CEE. 

The most significant difference between the two simulations is the speed of the inspiral phase. As seen in Figure~\ref{fig:sep}, the rejuvenated inspiral is about twice as fast as in the non-rejuvenated case. This is likely due to the abrupt increase in density and BE at the surface of the rejuvenated donor. The top panel of Figure~\ref{fig:BE} shows that much more mass is contained in the surface layers of the rejuvenated model compared to the non-rejuvenated case as a result of past accretion phases. These more massive outer layers exert a stronger gravitational drag on the companion as it grazes the envelope of the donor, which causes a stronger orbital decay and thus a much faster inspiral phase.

On the other hand, the difference in unbinding efficiency (the fraction of envelope mass that is unbound) between the two donor models is less substantial. From Figure~\ref{fig:mloss}, we do find an decrease in mass loss in the rejuvenated models when considering either energy criterion. However, the final contribution of internal energy to envelope unbinding is not certain at present, since we do not resolve the long-term evolution of the system and do not include radiative cooling or recombination energy, which may also influence the unbinding efficiency \citep[e.g.,][]{ivanova.2015,grichener.2018,soker.2018,lau_common_2022}. It is thus unclear whether the non-rejuvenated CEE is actually more efficient than the rejuvenated case. 

Overall, these results are severely limited by our resolution. Firstly, the largest difference in BE profiles in the two stellar models is situated in the innermost layers of the envelope ($r\sim1-10\,R_\odot$, see Figure~\ref{fig:BE}), which are not well-resolved with our current best resolution. Thus, simulating the CEE with smaller core radius and minimum cell size is required to properly assess the impact of rejuvenation on envelope ejection. Increasing the resolution would result in a longer inspiral phase before the cores coalesce or stabilize, thus improving the prediction of the outcome of the inspiral. If the binary stabilizes, simulating the inspiral around layers close to the CEB may also allow us to define a boundary between the ejected material and the remaining core \citep[e.g.,][]{tauris_research_2001,ivanova_common_2011,kruckow_common-envelope_2016} and parameterize 1D evolution models of the stripped remnant \citep[e.g.,][]{vigna-gomez_stellar_2022}. Additionally, correctly resolving the CEE on longer timescales will likely lead to an increase in mass loss due to a late conversion of internal energy to kinetic energy. 

However, due to the nature of 3D hydrodynamics methods, our simulations probe the initial dynamical phase of the evolution much better than later possible phases, which are longer-lived \citep[e.g.,][]{meyer.1979,renzo.2021}. Thus, our simulations are most sensitive to differences in the outer layers of the donors we use, which are also influenced by e.g., rotation and opacity effects \citep[e.g.,][]{blaauw.1993,renzo_evolution_2021,miszuda.2021}. Since the effects of rejuvenation are most apparent in the inner layers of the donor, close to the CEB, it may be too deep to be accurately probed with 3D hydrodynamics simulations at present.

Finally, since the mass loss in the systems considered in this study is very low, it may be of interest to study the effect of rejuvenation in systems with binary parameters allowing larger unbinding efficiencies. For instance, considering a more equal mass ratio can drastically increase the amount of mass lost during the inspiral, in which case the effect of previous rejuvenation may be much stronger. We might therefore expect the effects of rejuvenation to be more pronounced for progenitors of binary black hole systems than for progenitors of double neutron star systems. However, black holes mergers have been recently proposed to be more likely to form through stable mass transfer rather than CEE \citep[e.g.,][]{Pavlovskii.2017,vandenheuvel.2017,marchant_role_2021,klencki.2022,vanson.2022}.

\subsection{Shape of the outflow}
As seen in Figures~\ref{fig:snap}, \ref{fig:az_av}, and \ref{fig:mpolar}, one of the differences between our rejuvenated and non-rejuvenated simulations is the shape of the ejecta. While the ejecta are far from spherically symmetric in both cases, they are significantly more equatorially concentrated in the rejuvenated models. The differences in the symmetries and compactness of the ejecta are likely due to the different speeds at which the companion spirals in. In particular, a slower inspiral gives more time for the outflow to spread further out, which explains why the envelope is much more extended at the end of the non-rejuvenated simulation.

The ejecta asymmetries do not directly impact the CEE; however, they do have observational consequences as the ejected envelope expands and cools to the proper conditions for dust condensation. Thus, the resulting dusty CSM around the remnant of the CEE, which should be observable at least in the infrared, has a significantly different geometry and mass distribution if rejuvenation is taken into account. This may also impact the observational signatures of later interactions with the CSM.

The middle region between 5 and 15~AU in our simulations appears to be occupied by a mainly pressure-supported circumbinary disk. Pressure-supported disks have been observed in other common envelope simulations with massive donors \citep[][M.\ Vetter, private communication]{lau_common_2022}, and the long-term evolution of such a disk is likely to have major effects on the final separation \citep{metzger_luminous_2022,wei_evolution_2023}. In particular, this will have important consequences for the rates of gravitational wave events and merger-related transients. At present we can conclude that our simulations support the formation of a disk and suggest that its properties may be sensitive to the rejuvenation status of the donor. 

Since the companion in our simulations is a neutron star, a merger may produce a Thorne-\.{Z}ytkow object, which should most frequently originate from binary systems that experienced mass transfer and rejuvenation \citep{nathaniel_population_2024}. Alternatively, the merger could lead to collapse into a black hole accompanied by a strong supernova (SN) explosion \citep{chevalier_neutrino-cooled_1996,fryer_helium_1998}, and the dense circumstellar medium resulting from the CEE ejection will interact with the SN explosion \citep[e.g.,][]{smith_explosions_2011,chevalier_common_2012,metzger_luminous_2022}, potentially yielding a Type IIn SN or an ultra-luminous SN. The different equatorial concentrations and asymmetries observed between the rejuvenated and non-rejuvenated cases will impact the interaction between the circumstellar medium and a future SN, which may in turn affect the observed properties of the interacting SN and the remnant.

\subsection{Other effects of binary evolution}
Our simulations indicate that considering one previous phase of mass transfer significantly impacts CEE. However, there are other phases of binary evolution that may substantially impact the structure of a future CEE donor and thus the details of the CEE process. In particular, we found that the change in the mass distribution in the outer layers of the donor accelerated the inspiral phase; thus, any process that affects these regions may be relevant.

\begin{figure}
    \centering
    \includegraphics[width=1\columnwidth]{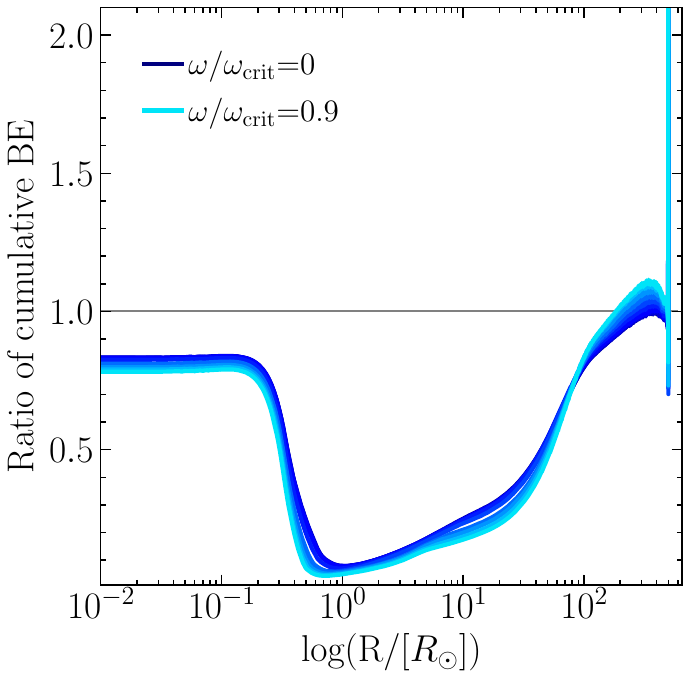}
    \caption{Ratio of cumulative BE of the 17.41~$M_\odot$ rejuvenated donor and 17.84~$M_\odot$ non-rejuvenated donors with different initial rotation rates, from $\omega/\omega_{\rm crit}=0$ (dark blue) to $\omega/\omega_{\rm crit}=0.9$ (cyan). Both models have a stellar radius of  500\,$R_\odot$. In our 3D simulations, we used the non-rejuvenated model with $\omega/\omega_{\rm crit}=0$, so the dark blue line corresponds to the black line in Fig~\ref{fig:BE}. The stellar models are from \cite{renzo_rejuvenated_2023}; see also their Figure~8.}
    \label{fig:BErot}
\end{figure}

Among these processes, stellar rotation is an important source of structural difference in a star. Beyond causing significant mixing, rotation impacts the size of the core \citep{heger.2000,maeder.2000} and the ensuing centrifugal force can drive or strengthen mass loss \citep{Langer.1998}. While a star can be born and evolve with an already significant rotation rate, it may also gain angular momentum through accretion \citep[e.g.,][]{packet_spin-up_1981,renzo_evolution_2021}, in which case the consequence of the spin-up should differ from the effect of natal rotation \citep{renzo_evolution_2021}. In this work, we considered a donor that has been spun up due to accretion, but compared it with a non-rotating and non-rejuvenated single donor, which is the type of donor profile most commonly used in 3D hydrodynamics simulations of CEE. This may, however, lead to different results than if we had considered a rotating non-rejuvenated star. In Figure~\ref{fig:BErot}, we show the ratio of cumulative BE for a rejuvenated accretor and single star, for different rotation rates for the single star (from $\omega/\omega_{\rm crit}=0$ to $\omega/\omega_{\rm crit}=0.9$, using the models computed by \cite{renzo_rejuvenated_2023} (see their Appendix~A). From this figure, it is clear that the difference in BE profiles is not significantly affected by rotation in the innermost layers of the star. However, we observe a significant difference in the BE ratio in the outer layers, suggesting that rotation does indeed play a role in unbinding these layers, requiring more attentive investigation in any follow-up work.

Besides accretion, stars in close binaries can gain or lose angular momentum through equilibrium tides \citep[e.g.][]{zahn.1977,hut.1981,Ma.2023}, which, for example, cause the circularization and synchronization of the orbit. However, tidal interactions are notoriously complicated and uncertain, and non equilibrium effects such as dynamical tidal waves may further influence the state of the binary close to the onset of CEE \citep{MacLeod.2019}. Thus, it can be argued that using uncertain rotation rates predicted by theories of tidal interaction does not necessarily improve the accuracy of 3D simulations. It should then be enough to accurately account for the spin-up due to previous binary interactions and its effect on the structure of the star, and arbitrarily set the binary and the envelope close to co-rotation at the start of the CEE simulation.

In this study, we considered a system that evolved through a typical double NS binary formation channel \citep[e.g.,][]{tauris_formation_2017,vigna-gomez_formation_2018}. This evolutionary process starts with two massive MS stars in a close orbit, such that the system undergoes a phase of stable mass transfer when the more massive star evolves out of its MS phase and expands. The donor is gradually stripped of its mass while the MS accretor gains mass and undergoes a rejuvenation process. Eventually, the stripped donor collapses and turns into a NS without disrupting the binary \citep{renzo_massive_2019}. The initially less massive star later expands when leaving its MS phase, eventually leading to a second phase of mass transfer that becomes unstable and leads to CEE. This second mass transfer phase can be short-lived, but it may also occur on the thermal timescale of the accretor \citep[e.g.,][]{nandez_recombination_2015,pejcha.2017,blagorodnova_luminous_2021}  and thus significantly affect the structure of the donor.

While the first phase of mass transfer is accounted for in the stellar models used in this work \citep{renzo_rejuvenated_2023}, the effects of the SN explosion and the second phase of mass transfer have been neglected. In particular, the ejecta due to the SN explosion can interact with the companion \citep[e.g.,][]{hirai_comprehensive_2018, Ogata_observability_2021}, injecting energy into its outer envelope, which would cause expansion and possibly ablate part of the surface layers. Additionally, the phase of mass transfer directly preceding the onset of CEE should also affect the donor, in particular through the removal of its outer layers. These two processes may remove some of the dense outer layers of the donor that formed through the first phase of mass transfer, thus mitigating the increase in inspiral pace while maintaining a lower BE profile across the stellar envelope. Including these two other phases of the evolution of the binary may therefore yield different results than we observed in our simulations. This scenario may be further complicated by assuming that the orbit is eccentric at the onset of CEE \citep{vigna-gomez_common_2020}, which may delay the loss of orbital stability since eccentric mass transfer only happens during periastron passages during its early stages \citep[e.g.,][]{staff_hydrodynamic_2016}.

We note that due to computational restrictions, most of the effects mentioned here can only be included through the use of 1D stellar evolution codes. This means that the use of the resulting stellar profiles for the initial conditions of 3D simulations of CEE will be suffering from the notorious issues of the 1D approximation. In particular, 1D schemes are unable to correctly approximate some 3D processes, such as convective motion, which renders their estimation of the thermal structure in the outermost layers of red supergiants inherently inaccurate. While 3D methods are needed to capture such effects \citep[e.g,][]{chiavassa.2015,jiang.2015,jiang.2018,Goldberg.2022}, they are too expensive for this purpose, and the use of 1D simulations is necessary for now, despite the fact that their inaccuracies may mitigate the effects of binary interactions on the outcome of CEE.
\section{Conclusions}
\label{sec:conclusions}
In this work, we have performed 3D hydrodynamics simulations of CEE using donor models that take or do not take into account a previous phase of mass transfer. This mass transfer phase induced a rejuvenation process in the future donor of the CEE phase, which significantly lowered its binding energy in the innermost envelope. We compare the outcomes of the CEE process in our simulations and found that rejuvenation does impact the speed of the inspiral, the shape of the ejecta, and the amount of mass lost by the system during the inspiral.

The main difference observed when a past phase of mass transfer is accounted for is a twice-faster inspiral phase. This is likely due to the denser outer layers of the donor resulting from past accretion, which strengthens the drag on the companion when it starts to plunge into the envelope. In both rejuvenated and non-rejuvenated cases, we found that the orbit of the binary does not stabilize before the numerical cores coalesce, though this may change when increasing the resolution of the simulations or considering more massive companions.

On the other hand, the impact of donor rejuvenation on the mass loss of the system is less clear. We found a 20$\%$ decrease in unbound mass with rejuvenation when using the conservative energy criterion, and a $40\%$ decrease when including the internal energy of the gas. The unbinding efficiency of the system is overall very low, with up to $5\%$ and $7\%$ of the envelope mass ejected in the rejuvenated and non-rejuvenated cases, respectively. Therefore, it is complicated to assess the full impact of rejuvenation on mass ejection, and simulation of different mass ratios may help in that regard.

One effect of using rejuvenated stellar models was a change in ejecta morphology: the rejuvenated ejecta are more equatorially concentrated and less spherically symmetrical than in our non-rejuvenated models. This difference mostly has observational consequences, especially if the inspiral ends in a merger. In particular, the ensuing merger transient and a potential later SN explosion of the merger product may have different properties if the mass distribution of the CEE ejecta is different. In both rejuvenated and non-rejuvenated cases we observed the formation of a pressure-supported disk with different properties.

We note that due to numerical limitations, our simulations have relatively large core sizes that do not allow us to resolve the innermost part of the inspiral, where the difference in BE should be the strongest. As a result, both sets of simulations are consistent with either a merger of the cores or stabilization of the binary below our resolution limit. Therefore, simulations with smaller core sizes and higher resolution are needed to study the full impact of rejuvenation.

Our simulations only considered a specific phase of the typical binary evolution channel for double NS binaries, but the results indicate that any significant structural changes in the stars occurring during binary interactions are relevant for later CEE. Thus, it seems important for future CEE simulations to account for as much of the binary evolution previous to the onset of CEE as possible, especially previous mass transfer phases and structural changes due to rotation.

\begin{acknowledgments}
This project was initiated at the 2023 Kavli Summer Program in Astrophysics at the Max Planck Institute for Astrophysics (MPA). We thank MPA and the Kavli Program for their support. 
CL has been supported by Horizon 2020 ERC Starting Grant ‘Cat-In-hAT’ (grant agreement no.\ 803158) and the Charles University Grant Agency project no.\ 116324. This work was also supported by the Ministry of Education, Youth and Sports of the Czech Republic through the e-INFRA CZ (ID:90254). PR and SR acknowledge support from the US National Science Foundation under AAG 23-07713.
AVG acknowledges funding from the Netherlands Organisation for Scientific Research (NWO), as part of the Vidi research program BinWaves (project number 639.042.728, PI: de Mink).
FLASH is maintained largely by the DOE-supported Flash Center for Computational Science at the University of Rochester. 
Simulations were performed using Barbora at IT4Innovations (projects OPEN-27-60 and OPEN-30-50), Stampede2 at the Texas Advanced Computing Center (NSF ACCESS PHY230145), and machines at MPA and Charles University.
\end{acknowledgments}

\software{MESA \citep{paxton_modules_2011, paxton_modules_2013, paxton_modules_2015, paxton_modules_2018, paxton_modules_2019},
          FLASH \citep{fryxell_flash_2000,dubey_introduction_2008},
          yt \citep{turk_yt_2011},
          Matplotlib \citep{hunter_matplotlib_2007},
          NumPy \citep{harris_array_2020}
          }


\appendix
\section{Level of error in conserved quantities}

We tracked changes in total quantities that are not explicitly conserved by FLASH or that can change because of our diode boundary conditions and report those results for the highest-resolution runs here. We include all gas represented on the grid (including the low-density fluff, whose initial mass was about $0.6M_\odot$) as well as the stellar cores in these quantities.

\begin{figure}
    \centering
    \includegraphics[height=2.5in]{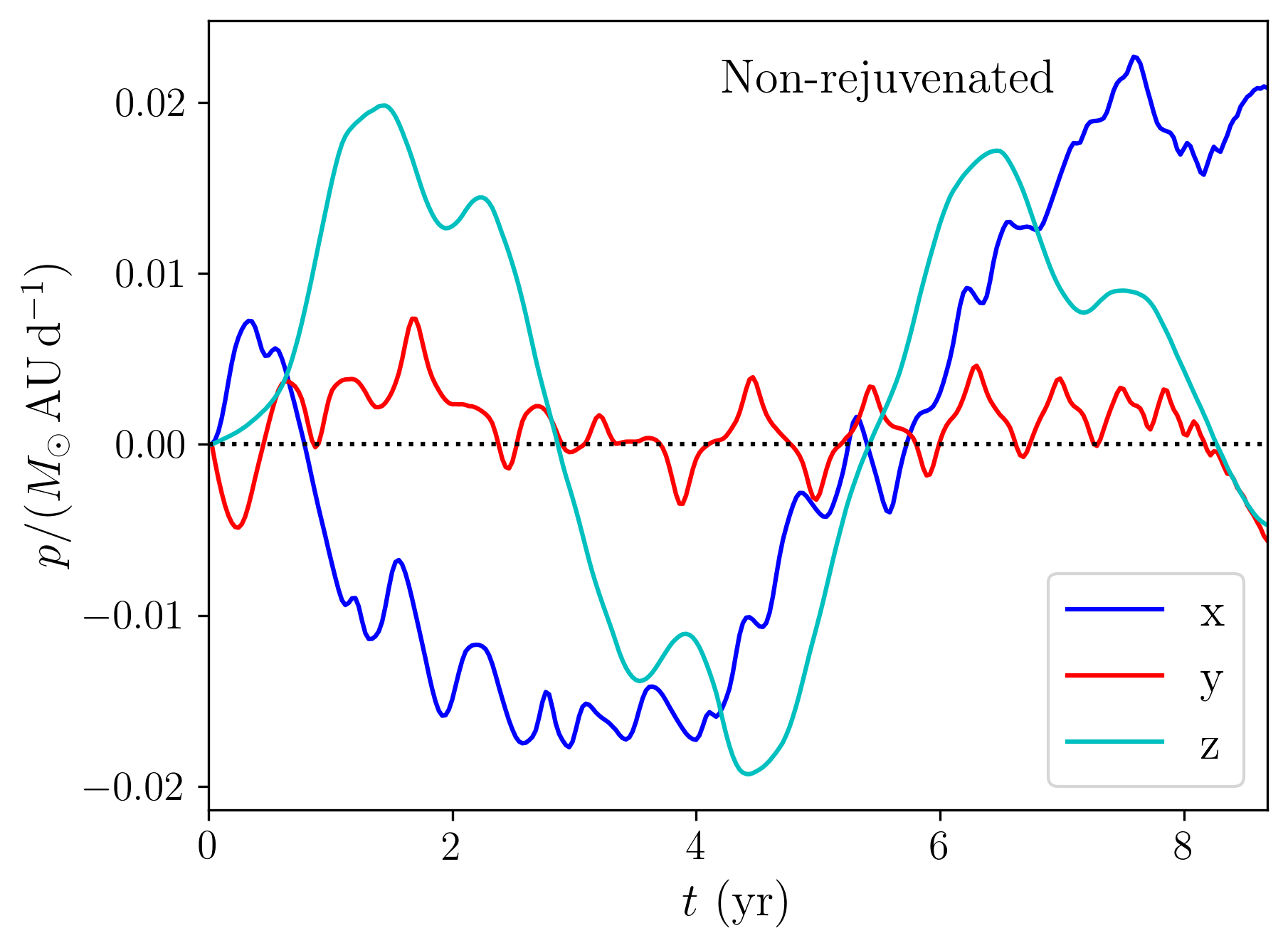}
    \includegraphics[height=2.5in]{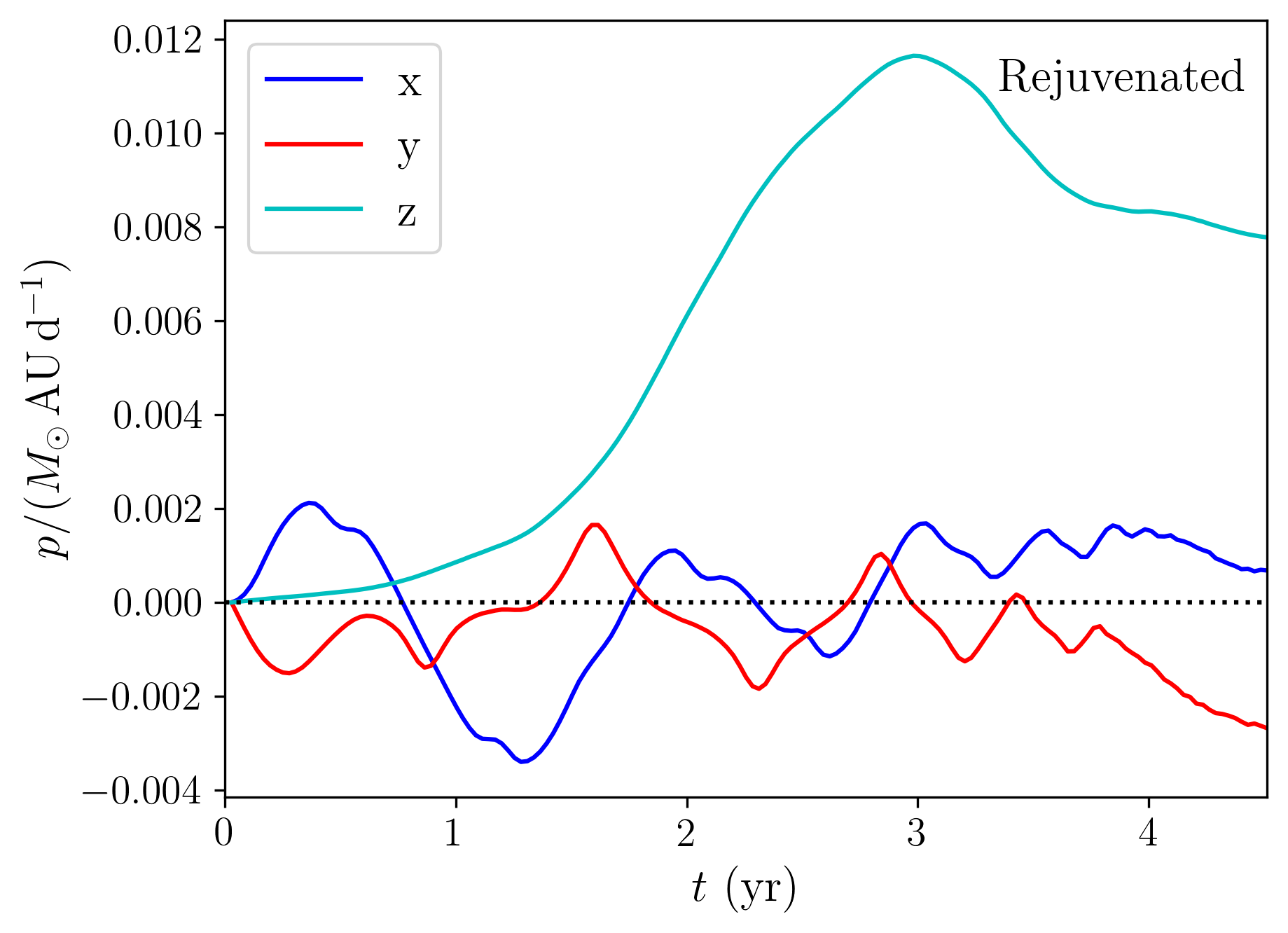}
    \caption{Evolution of $x$, $y$, and $z$ components of the total momentum. Left:
             \nonrej(12, 1.26). Right: \rej(12, 1.26).}
    \label{fig:linmom}
\end{figure}

The evolution of the linear momentum is shown in Figure~\ref{fig:linmom}. Because the initial linear momentum is zero, we plot the absolute rather than relative change. Its behavior is oscillatory, with the largest amplitudes for the $z$~component in both cases and the $x$~component for the non-rejuvenated case. The net drift in the center of mass by the end of the simulation is $[0.013, -0.070, 0.47]\ {\rm AU}$ for the rejuvenated case and $[-0.20, -0.025, 0.52]\ {\rm AU}$ for the non-rejuvenated case.

\begin{figure}
    \centering
    \includegraphics[height=2.5in]{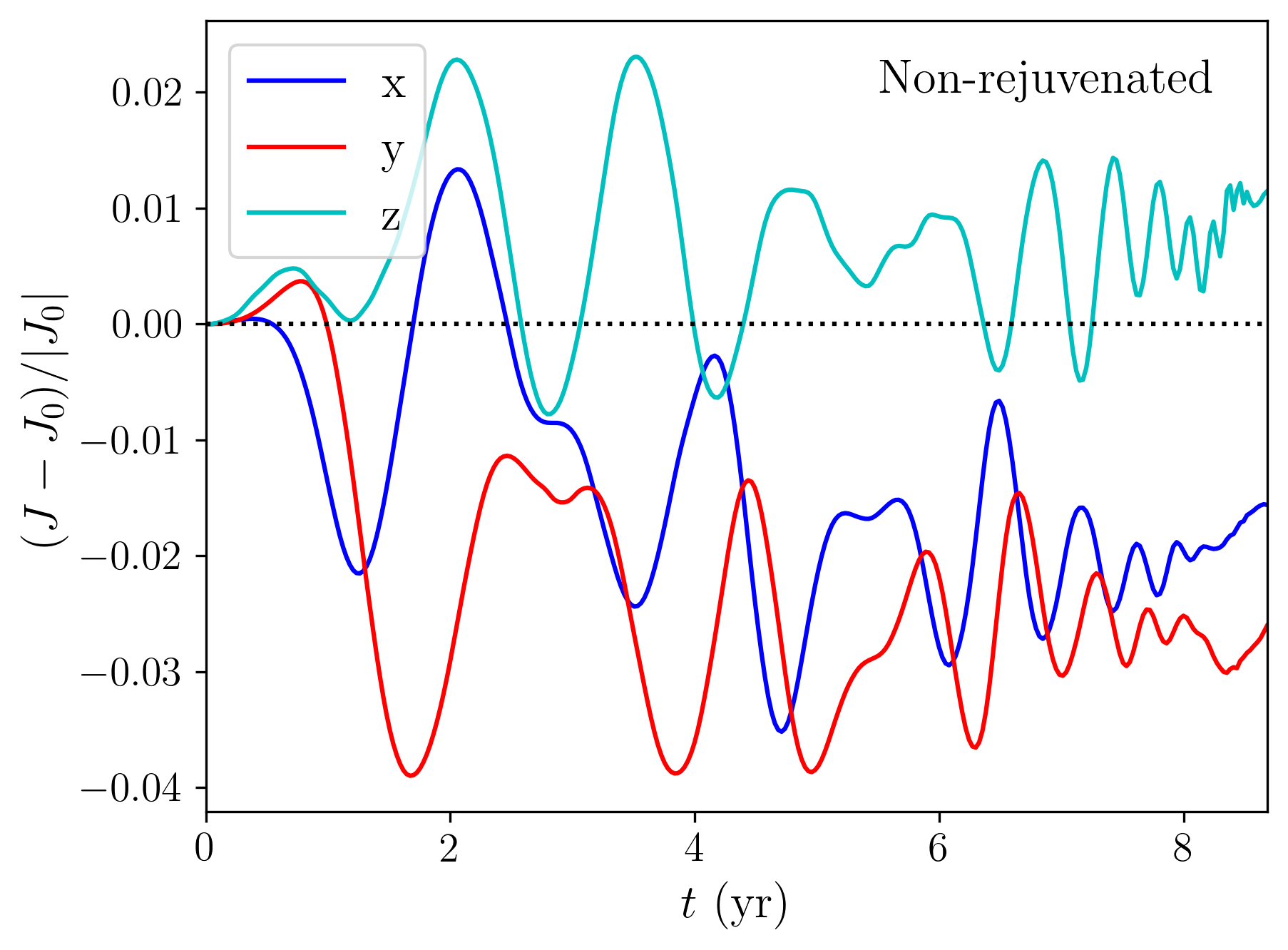}
    \includegraphics[height=2.5in]{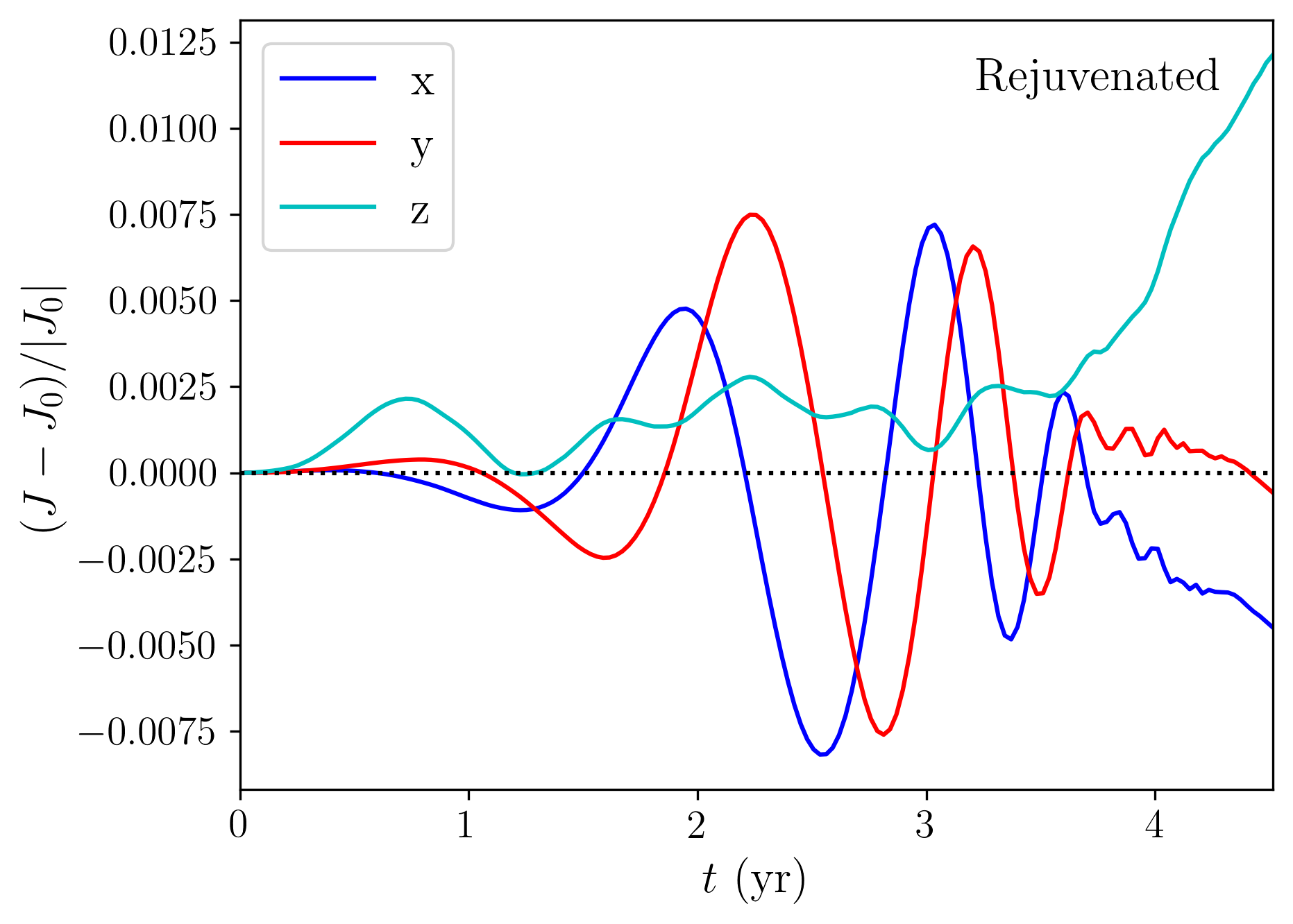}
    \caption{Evolution of $x$, $y$, and $z$ components of the total angular momentum. Left:
             \nonrej(12, 1.26). Right: \rej(12, 1.26).}
    \label{fig:angmom}
\end{figure}

The evolution of the fractional change in angular momentum is shown in Figure~\ref{fig:angmom}. The angular momentum about the center of mass is dominated by the $z$~component, which was initially $0.26\ M_\odot\ {\rm AU}^2\ {\rm d}^{-1}$ for the rejuvenated case and $0.23\ M_\odot\ {\rm AU}^2\ {\rm d}^{-1}$ for the non-rejuvenated case. All components change by less than 4\% of these initial values over the course of the runs, with the $z$~component changing by less than 2.3\%. As the core and companion particles inspiral from an initial separation of about 4~AU to a minimum resolvable separation of about 0.12~AU, their angular momentum decreases by a factor of 5.9 (i.e.,\ to 17\% of its initial value), so non-conservation of angular momentum may introduce as much as a 27\% error in the final separation.

\begin{figure}
    \centering
    \includegraphics[height=2.5in]{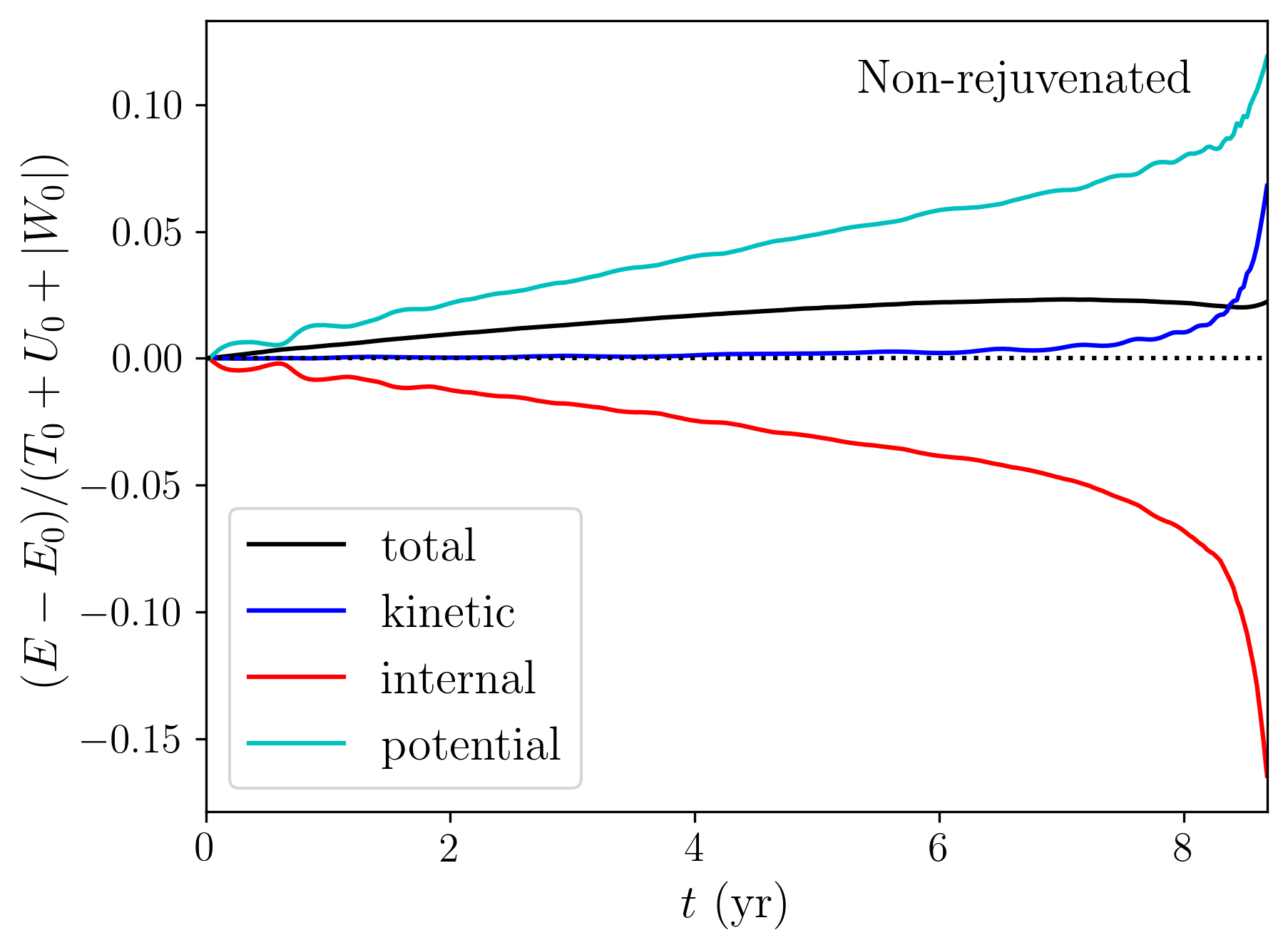}
    \includegraphics[height=2.5in]{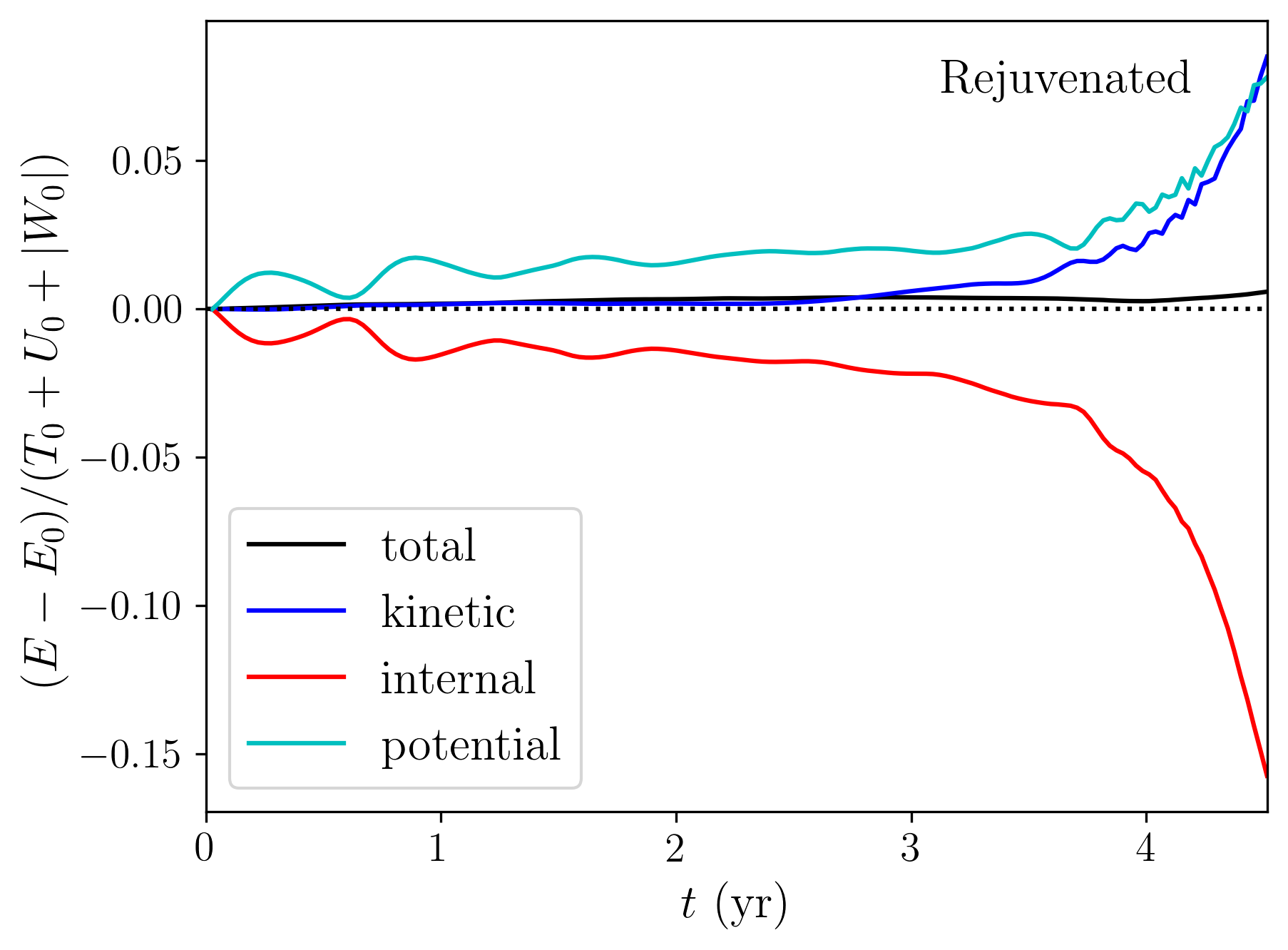}
    \caption{Evolution of the total energy and its constituents relative to the initial sum
             of the kinetic and internal energies and the absolute value of the potential
             energy. Left:
             \nonrej(12, 1.26). Right: \rej(12, 1.26).}
    \label{fig:energy}
\end{figure}

The evolution of the fractional change in total energy and its constituents appears in Figure~\ref{fig:energy}. Very little ($< 0.1\%$) of the envelope mass escapes the grid during either simulation. Hence the change in total energy is a good measure of overall energy conservation. We find that in the rejuvenated case, the total energy changes by less than 1\% of the sum of the kinetic and internal energies and the absolute value of the potential energy, while for the non-rejuvenated case the total change is less than 2.5\%.


\bibliography{ms}{}

\begin{thebibliography}{}
\expandafter\ifx\csname natexlab\endcsname\relax\def\natexlab#1{#1}\fi
\providecommand{\url}[1]{\href{#1}{#1}}
\providecommand{\dodoi}[1]{doi:~\href{http://doi.org/#1}{\nolinkurl{#1}}}
\providecommand{\doeprint}[1]{\href{http://ascl.net/#1}{\nolinkurl{http://ascl.net/#1}}}
\providecommand{\doarXiv}[1]{\href{https://arxiv.org/abs/#1}{\nolinkurl{https://arxiv.org/abs/#1}}}

\bibitem[{Abt \& Levy(1976)}]{abt_multiplicity_1976}
Abt, H.~A., \& Levy, S.~G. 1976, \apjs, 30, 273, \dodoi{10.1086/190363}

\bibitem[{Belczynski {et~al.}(2016)Belczynski, Holz, Bulik, \& O'Shaughnessy}]{belczynski_first_2016}
Belczynski, K., Holz, D.~E., Bulik, T., \& O'Shaughnessy, R. 2016, \nat, 534, 512, \dodoi{10.1038/nature18322}

\bibitem[{{Blaauw}(1993)}]{blaauw.1993}
{Blaauw}, A. 1993, in Astronomical Society of the Pacific Conference Series, Vol.~35, Massive Stars: Their Lives in the Interstellar Medium, ed. J.~P. {Cassinelli} \& E.~B. {Churchwell}, 207

\bibitem[{Blagorodnova {et~al.}(2021)Blagorodnova, Klencki, Pejcha, Vreeswijk, Bond, Burdge, De, Fremling, Gehrz, Jencson, Kasliwal, Kupfer, Lau, Masci, \& Rich}]{blagorodnova_luminous_2021}
Blagorodnova, N., Klencki, J., Pejcha, O., {et~al.} 2021, \aap, 653, A134, \dodoi{10.1051/0004-6361/202140525}

\bibitem[{Bonnell {et~al.}(2004)Bonnell, Vine, \& Bate}]{bonnell_massive_2004}
Bonnell, I.~A., Vine, S.~G., \& Bate, M.~R. 2004, \mnras, 349, 735, \dodoi{10.1111/j.1365-2966.2004.07543.x}

\bibitem[{Cantiello {et~al.}(2007)Cantiello, Yoon, Langer, \& Livio}]{cantiello_binary_2007}
Cantiello, M., Yoon, S.~C., Langer, N., \& Livio, M. 2007, \aap, 465, L29, \dodoi{10.1051/0004-6361:20077115}

\bibitem[{Chevalier(1996)}]{chevalier_neutrino-cooled_1996}
Chevalier, R.~A. 1996, \apj, 459, 322, \dodoi{10.1086/176895}

\bibitem[{Chevalier(2012)}]{chevalier_common_2012}
---. 2012, \apj, 752, L2, \dodoi{10.1088/2041-8205/752/1/L2}

\bibitem[{{Chiavassa} \& {Freytag}(2015)}]{chiavassa.2015}
{Chiavassa}, A., \& {Freytag}, B. 2015, in Astronomical Society of the Pacific Conference Series, Vol. 497, Why Galaxies Care about AGB Stars III: A Closer Look in Space and Time, ed. F.~{Kerschbaum}, R.~F. {Wing}, \& J.~{Hron}, 11, \dodoi{10.48550/arXiv.1410.3868}

\bibitem[{Daley {et~al.}(2012)Daley, Vanella, Dubey, Weide, \& Balaras}]{daley_optimization_2012}
Daley, C., Vanella, M., Dubey, A., Weide, K., \& Balaras, E. 2012, Concurrency and Computation: Practice and Experience, 24, 2346, \dodoi{10.1002/cpe}

\bibitem[{De~Marco \& Izzard(2017)}]{de_marco_dawes_2017}
De~Marco, O., \& Izzard, R.~G. 2017, \pasa, 34, e001, \dodoi{10.1017/pasa.2016.52}

\bibitem[{Dubey {et~al.}(2008)Dubey, Reid, \& Fisher}]{dubey_introduction_2008}
Dubey, A., Reid, L.~B., \& Fisher, R. 2008, PhyS, T132, 014046, \dodoi{10.1088/0031-8949/2008/T132/014046}

\bibitem[{Duchêne \& Kraus(2013)}]{duchene_stellar_2013}
Duchêne, G., \& Kraus, A. 2013, \araa, 51, 269, \dodoi{10.1146/annurev-astro-081710-102602}

\bibitem[{Eggleton(1983)}]{eggleton_approximations_1983}
Eggleton, P.~P. 1983, \apj, 268, 368, \dodoi{10.1086/160960}

\bibitem[{Emden(1907)}]{emden_gaskugeln_1907}
Emden, R. 1907, Gaskugeln: {Anwendungen} der mechanischen {Wärmetheorie} auf kosmologische und meteorologische {Probleme} (Leipzig: B.G. Teubner)

\bibitem[{Fryer \& Woosley(1998)}]{fryer_helium_1998}
Fryer, C.~L., \& Woosley, S.~E. 1998, \apjl, 502, L9, \dodoi{10.1086/311493}

\bibitem[{Fryxell {et~al.}(2000)Fryxell, Olson, Ricker, Timmes, Zingale, Lamb, MacNeice, Rosner, Truran, \& Tufo}]{fryxell_flash_2000}
Fryxell, B., Olson, K., Ricker, P., {et~al.} 2000, \apjs, 131, 273, \dodoi{10.1086/317361}

\bibitem[{{Goldberg} {et~al.}(2022){Goldberg}, {Jiang}, \& {Bildsten}}]{Goldberg.2022}
{Goldberg}, J.~A., {Jiang}, Y.-F., \& {Bildsten}, L. 2022, \apj, 929, 156, \dodoi{10.3847/1538-4357/ac5ab3}

\bibitem[{{Grichener} {et~al.}(2018){Grichener}, {Sabach}, \& {Soker}}]{grichener.2018}
{Grichener}, A., {Sabach}, E., \& {Soker}, N. 2018, \mnras, 478, 1818, \dodoi{10.1093/mnras/sty1178}

\bibitem[{Harris {et~al.}(2020)Harris, Millman, Walt, Gommers, Virtanen, Cournapeau, Wieser, Taylor, Berg, Smith, Kern, Picus, Hoyer, Kerkwijk, Brett, Haldane, Río, Wiebe, Peterson, Gérard-Marchant, Sheppard, Reddy, Weckesser, Abbasi, Gohlke, \& Oliphant}]{harris_array_2020}
Harris, C.~R., Millman, K.~J., Walt, S. J. v.~d., {et~al.} 2020, \nat, 585, 357, \dodoi{10.1038/s41586-020-2649-2}

\bibitem[{{Heger} {et~al.}(2000){Heger}, {Langer}, \& {Woosley}}]{heger.2000}
{Heger}, A., {Langer}, N., \& {Woosley}, S.~E. 2000, \apj, 528, 368, \dodoi{10.1086/308158}

\bibitem[{Hellings(1983)}]{hellings_phenomenological_1983}
Hellings, P. 1983, \apss, 96, 37, \dodoi{10.1007/BF00661941}

\bibitem[{Hirai {et~al.}(2018)Hirai, Podsiadlowski, \& Yamada}]{hirai_comprehensive_2018}
Hirai, R., Podsiadlowski, P., \& Yamada, S. 2018, \apj, 864, 119, \dodoi{10.3847/1538-4357/aad6a0}

\bibitem[{Hunter(2007)}]{hunter_matplotlib_2007}
Hunter, J.~D. 2007, CSE, 9, 90, \dodoi{10.1109/MCSE.2007.55}

\bibitem[{{Hut}(1981)}]{hut.1981}
{Hut}, P. 1981, \aap, 99, 126

\bibitem[{Ivanova(2011)}]{ivanova_common_2011}
Ivanova, N. 2011, \apj, 730, 76, \dodoi{10.1088/0004-637X/730/2/76}

\bibitem[{Ivanova {et~al.}(2013{\natexlab{a}})Ivanova, Justham, Nandez, \& Lombardi~Jr}]{ivanova_identification_2013}
Ivanova, N., Justham, S., Nandez, J. L.~A., \& Lombardi~Jr, J.~C. 2013{\natexlab{a}}, Science, 339, 433, \dodoi{10.1126/science.1225540}

\bibitem[{{Ivanova} {et~al.}(2015){Ivanova}, {Justham}, \& {Podsiadlowski}}]{ivanova.2015}
{Ivanova}, N., {Justham}, S., \& {Podsiadlowski}, P. 2015, \mnras, 447, 2181, \dodoi{10.1093/mnras/stu2582}

\bibitem[{Ivanova {et~al.}(2020)Ivanova, Justham, \& Ricker}]{ivanova_common_2020}
Ivanova, N., Justham, S., \& Ricker, P. 2020, Common {Envelope} {Evolution}, {AAS}-{IOP} {Astronomy} ebooks (IOP Publishing).
\newblock \url{https://iopscience.iop.org/book/978-0-7503-1563-0}

\bibitem[{Ivanova {et~al.}(2013{\natexlab{b}})Ivanova, Justham, Chen, De~Marco, Fryer, Gaburov, Ge, Glebbeek, Han, Li, Lu, Marsh, Podsiadlowski, Potter, Soker, Taam, Tauris, van~den Heuvel, \& Webbink}]{ivanova_common_2013}
Ivanova, N., Justham, S., Chen, X., {et~al.} 2013{\natexlab{b}}, \aapr, 21, 59, \dodoi{10.1007/s00159-013-0059-2}

\bibitem[{{Jiang} {et~al.}(2015){Jiang}, {Cantiello}, {Bildsten}, {Quataert}, \& {Blaes}}]{jiang.2015}
{Jiang}, Y.-F., {Cantiello}, M., {Bildsten}, L., {Quataert}, E., \& {Blaes}, O. 2015, \apj, 813, 74, \dodoi{10.1088/0004-637X/813/1/74}

\bibitem[{{Jiang} {et~al.}(2018){Jiang}, {Cantiello}, {Bildsten}, {Quataert}, {Blaes}, \& {Stone}}]{jiang.2018}
{Jiang}, Y.-F., {Cantiello}, M., {Bildsten}, L., {et~al.} 2018, \nat, 561, 498, \dodoi{10.1038/s41586-018-0525-0}

\bibitem[{Kalogera \& Webbink(1998)}]{Kalogera_formation_1998}
Kalogera, V., \& Webbink, R.~F. 1998, \apj, 493, 351, \dodoi{10.1086/305085}

\bibitem[{Kashi \& Soker(2011)}]{kashi_circumbinary_2011}
Kashi, A., \& Soker, N. 2011, \mnras, 417, 1466, \dodoi{10.1111/j.1365-2966.2011.19361.x}

\bibitem[{{Klencki} {et~al.}(2022){Klencki}, {Istrate}, {Nelemans}, \& {Pols}}]{klencki.2022}
{Klencki}, J., {Istrate}, A., {Nelemans}, G., \& {Pols}, O. 2022, \aap, 662, A56, \dodoi{10.1051/0004-6361/202142701}

\bibitem[{Klencki {et~al.}(2021)Klencki, Nelemans, Istrate, \& Chruslinska}]{klencki_it_2021}
Klencki, J., Nelemans, G., Istrate, A.~G., \& Chruslinska, M. 2021, \aap, 645, A54, \dodoi{10.1051/0004-6361/202038707}

\bibitem[{Kruckow {et~al.}(2016)Kruckow, Tauris, Langer, Szécsi, Marchant, \& Podsiadlowski}]{kruckow_common-envelope_2016}
Kruckow, M.~U., Tauris, T.~M., Langer, N., {et~al.} 2016, \aap, 596, A58, \dodoi{10.1051/0004-6361/201629420}

\bibitem[{Lane(1870)}]{lane_theoretical_1870}
Lane, H.~J. 1870, AmJS, 50, 57, \dodoi{10.2475/ajs.s2-50.148.57}

\bibitem[{{Langer}(1998)}]{Langer.1998}
{Langer}, N. 1998, \aap, 329, 551

\bibitem[{Langer(2012)}]{langer_presupernova_2012}
Langer, N. 2012, \araa, 50, 107, \dodoi{10.1146/annurev-astro-081811-125534}

\bibitem[{Lau {et~al.}(2022)Lau, Hirai, González-Bolívar, Price, De~Marco, \& Mandel}]{lau_common_2022}
Lau, M. Y.~M., Hirai, R., González-Bolívar, M., {et~al.} 2022, \mnras, 512, 5462, \dodoi{10.1093/mnras/stac049}

\bibitem[{{Law-Smith} {et~al.}(2020){Law-Smith}, {Everson}, {Ramirez-Ruiz}, {de Mink}, {van Son}, {G{\"o}tberg}, {Zellmann}, {Vigna-G{\'o}mez}, {Renzo}, {Wu}, {Schr{\o}der}, {Foley}, \& {Hutchinson-Smith}}]{lawsmith.2020}
{Law-Smith}, J. A.~P., {Everson}, R.~W., {Ramirez-Ruiz}, E., {et~al.} 2020, arXiv e-prints, \dodoi{10.48550/arXiv.2011.06630}

\bibitem[{{Ma} \& {Fuller}(2023)}]{Ma.2023}
{Ma}, L., \& {Fuller}, J. 2023, \apj, 952, 53, \dodoi{10.3847/1538-4357/acdb74}

\bibitem[{MacLeod {et~al.}(2018)MacLeod, Ostriker, \& Stone}]{macleod_bound_2018}
MacLeod, M., Ostriker, E.~C., \& Stone, J.~M. 2018, \apj, 868, 136, \dodoi{10.3847/1538-4357/aae9eb}

\bibitem[{{MacLeod} {et~al.}(2019){MacLeod}, {Vick}, {Lai}, \& {Stone}}]{MacLeod.2019}
{MacLeod}, M., {Vick}, M., {Lai}, D., \& {Stone}, J.~M. 2019, \apj, 877, 28, \dodoi{10.3847/1538-4357/ab184c}

\bibitem[{{Maeder} \& {Meynet}(2000)}]{maeder.2000}
{Maeder}, A., \& {Meynet}, G. 2000, \araa, 38, 143, \dodoi{10.1146/annurev.astro.38.1.143}

\bibitem[{Marchant {et~al.}(2021)Marchant, Pappas, Gallegos-Garcia, Berry, Taam, Kalogera, \& Podsiadlowski}]{marchant_role_2021}
Marchant, P., Pappas, K. M.~W., Gallegos-Garcia, M., {et~al.} 2021, \aap, 650, A107, \dodoi{10.1051/0004-6361/202039992}

\bibitem[{Mason {et~al.}(2009)Mason, Hartkopf, Gies, Henry, \& Helsel}]{mason_high_2009}
Mason, B.~D., Hartkopf, W.~I., Gies, D.~R., Henry, T.~J., \& Helsel, J.~W. 2009, \aj, 137, 3358, \dodoi{10.1088/0004-6256/137/2/3358}

\bibitem[{Metzger(2022)}]{metzger_luminous_2022}
Metzger, B.~D. 2022, \apj, 932, 84, \dodoi{10.3847/1538-4357/ac6d59}

\bibitem[{{Meyer} \& {Meyer-Hofmeister}(1979)}]{meyer.1979}
{Meyer}, F., \& {Meyer-Hofmeister}, E. 1979, \aap, 78, 167

\bibitem[{{Miszuda} {et~al.}(2021){Miszuda}, {Szewczuk}, \& {Daszy{\'n}ska-Daszkiewicz}}]{miszuda.2021}
{Miszuda}, A., {Szewczuk}, W., \& {Daszy{\'n}ska-Daszkiewicz}, J. 2021, \mnras, 505, 3206, \dodoi{10.1093/mnras/stab1597}

\bibitem[{Moe \& Di~Stefano(2017)}]{moe_mind_2017}
Moe, M., \& Di~Stefano, R. 2017, \apjs, 230, 15, \dodoi{10.3847/1538-4365/aa6fb6}

\bibitem[{Moreno {et~al.}(2022)Moreno, Schneider, Röpke, Ohlmann, Pakmor, Podsiadlowski, \& Sand}]{moreno_3d_2022}
Moreno, M.~M., Schneider, F. R.~N., Röpke, F.~K., {et~al.} 2022, \aap, 667, A72, \dodoi{10.1051/0004-6361/202142731}

\bibitem[{Nandez {et~al.}(2015)Nandez, Ivanova, \& Lombardi}]{nandez_recombination_2015}
Nandez, J. L.~A., Ivanova, N., \& Lombardi, J. C.~J. 2015, \mnras, 450, L39, \dodoi{10.1093/mnrasl/slv043}

\bibitem[{{Nathaniel} {et~al.}(2024){Nathaniel}, {Vigna-G{\'o}mez}, {Grichener}, {Farmer}, {Renzo}, \& {Everson}}]{nathaniel_population_2024}
{Nathaniel}, K., {Vigna-G{\'o}mez}, A., {Grichener}, A., {et~al.} 2024, arXiv e-prints.
\newblock \doarXiv{2407.11680}

\bibitem[{{Offner} {et~al.}(2023){Offner}, {Moe}, {Kratter}, {Sadavoy}, {Jensen}, \& {Tobin}}]{offner_origin_2023}
{Offner}, S.~S.~R., {Moe}, M., {Kratter}, K.~M., {et~al.} 2023, in Astronomical Society of the Pacific Conference Series, Vol. 534, Protostars and Planets VII, ed. S.~{Inutsuka}, Y.~{Aikawa}, T.~{Muto}, K.~{Tomida}, \& M.~{Tamura}, 275, \dodoi{10.48550/arXiv.2203.10066}

\bibitem[{Ogata {et~al.}(2021)Ogata, Hirai, \& Hijikawa}]{Ogata_observability_2021}
Ogata, M., Hirai, R., \& Hijikawa, K. 2021, \mnras, 505, 2485, \dodoi{10.1093/mnras/stab1439}

\bibitem[{Ohlmann {et~al.}(2016)Ohlmann, Röpke, Pakmor, \& Springel}]{ohlmann_hydrodynamic_2016}
Ohlmann, S.~T., Röpke, F.~K., Pakmor, R., \& Springel, V. 2016, \apj, 816, L9, \dodoi{10.3847/2041-8205/816/1/L9}

\bibitem[{Ohlmann {et~al.}(2017)Ohlmann, Röpke, Pakmor, \& Springel}]{ohlmann_constructing_2017}
---. 2017, \aap, 599, A5, \dodoi{10.1051/0004-6361/201629692}

\bibitem[{Packet(1981)}]{packet_spin-up_1981}
Packet, W. 1981, \aap, 102, 17.
\newblock \url{https://ui.adsabs.harvard.edu/abs/1981A&A...102...17P}

\bibitem[{{Paczynski}(1976)}]{paczynski_common_1976}
{Paczynski}, B. 1976, in Structure and Evolution of Close Binary Systems, ed. P.~{Eggleton}, S.~{Mitton}, \& J.~{Whelan}, Vol.~73, 75

\bibitem[{{Pavlovskii} {et~al.}(2017){Pavlovskii}, {Ivanova}, {Belczynski}, \& {Van}}]{Pavlovskii.2017}
{Pavlovskii}, K., {Ivanova}, N., {Belczynski}, K., \& {Van}, K.~X. 2017, \mnras, 465, 2092, \dodoi{10.1093/mnras/stw2786}

\bibitem[{Paxton {et~al.}(2011)Paxton, Bildsten, Dotter, Herwig, Lesaffre, \& Timmes}]{paxton_modules_2011}
Paxton, B., Bildsten, L., Dotter, A., {et~al.} 2011, \apjs, 192, 3, \dodoi{10.1088/0067-0049/192/1/3}

\bibitem[{Paxton {et~al.}(2013)Paxton, Cantiello, Arras, Bildsten, Brown, Dotter, Mankovich, Montgomery, Stello, Timmes, \& Townsend}]{paxton_modules_2013}
Paxton, B., Cantiello, M., Arras, P., {et~al.} 2013, \apjs, 208, 4, \dodoi{10.1088/0067-0049/208/1/4}

\bibitem[{Paxton {et~al.}(2015)Paxton, Marchant, Schwab, Bauer, Bildsten, Cantiello, Dessart, Farmer, Hu, Langer, Townsend, Townsley, \& Timmes}]{paxton_modules_2015}
Paxton, B., Marchant, P., Schwab, J., {et~al.} 2015, \apjs, 220, 15, \dodoi{10.1088/0067-0049/220/1/15}

\bibitem[{Paxton {et~al.}(2018)Paxton, Schwab, Bauer, Bildsten, Blinnikov, Duffell, Farmer, Goldberg, Marchant, Sorokina, Thoul, Townsend, \& Timmes}]{paxton_modules_2018}
Paxton, B., Schwab, J., Bauer, E.~B., {et~al.} 2018, \apjs, 234, 34, \dodoi{10.3847/1538-4365/aaa5a8}

\bibitem[{Paxton {et~al.}(2019)Paxton, Smolec, Schwab, Gautschy, Bildsten, Cantiello, Dotter, Farmer, Goldberg, Jermyn, Kanbur, Marchant, Thoul, Townsend, Wolf, Zhang, \& Timmes}]{paxton_modules_2019}
Paxton, B., Smolec, R., Schwab, J., {et~al.} 2019, \apjs, 243, 10, \dodoi{10.3847/1538-4365/ab2241}

\bibitem[{Pejcha {et~al.}(2016)Pejcha, Metzger, \& Tomida}]{pejcha_cool_2016}
Pejcha, O., Metzger, B.~D., \& Tomida, K. 2016, \mnras, 455, 4351, \dodoi{10.1093/mnras/stv2592}

\bibitem[{{Pejcha} {et~al.}(2017){Pejcha}, {Metzger}, {Tyles}, \& {Tomida}}]{pejcha.2017}
{Pejcha}, O., {Metzger}, B.~D., {Tyles}, J.~G., \& {Tomida}, K. 2017, \apj, 850, 59, \dodoi{10.3847/1538-4357/aa95b9}

\bibitem[{Podsiadlowski {et~al.}(1992)Podsiadlowski, Joss, \& Hsu}]{podsiadlowski_presupernova_1992}
Podsiadlowski, P., Joss, P.~C., \& Hsu, J. J.~L. 1992, \apj, 391, 246, \dodoi{10.1086/171341}

\bibitem[{Reichardt {et~al.}(2020)Reichardt, De~Marco, Iaconi, Chamandy, \& Price}]{reichardt_impact_2020}
Reichardt, T.~A., De~Marco, O., Iaconi, R., Chamandy, L., \& Price, D.~J. 2020, \mnras, 494, 5333, \dodoi{10.1093/mnras/staa937}

\bibitem[{Renzo \& Götberg(2021)}]{renzo_evolution_2021}
Renzo, M., \& Götberg, Y. 2021, \apj, 923, 277, \dodoi{10.3847/1538-4357/ac29c5}

\bibitem[{Renzo {et~al.}(2023)Renzo, Zapartas, Justham, Breivik, Lau, Farmer, Cantiello, \& Metzger}]{renzo_rejuvenated_2023}
Renzo, M., Zapartas, E., Justham, S., {et~al.} 2023, \apj, 942, L32, \dodoi{10.3847/2041-8213/aca4d3}

\bibitem[{Renzo {et~al.}(2019)Renzo, Zapartas, de~Mink, Götberg, Justham, Farmer, Izzard, Toonen, \& Sana}]{renzo_massive_2019}
Renzo, M., Zapartas, E., de~Mink, S.~E., {et~al.} 2019, \aap, 624, A66, \dodoi{10.1051/0004-6361/201833297}

\bibitem[{{Renzo} {et~al.}(2021){Renzo}, {Callister}, {Chatziioannou}, {van Son}, {Mingarelli}, {Cantiello}, {Ford}, {McKernan}, \& {Ashton}}]{renzo.2021}
{Renzo}, M., {Callister}, T., {Chatziioannou}, K., {et~al.} 2021, \apj, 919, 128, \dodoi{10.3847/1538-4357/ac1110}

\bibitem[{Ricker(2008)}]{ricker_direct_2008}
Ricker, P.~M. 2008, \apjs, 176, 293, \dodoi{10.1086/526425}

\bibitem[{{Ricker} \& {Taam}(2012)}]{ricker_taam_2012}
{Ricker}, P.~M., \& {Taam}, R.~E. 2012, \apj, 746, 74, \dodoi{10.1088/0004-637X/746/1/74}

\bibitem[{Röpke \& De~Marco(2023)}]{ropke_simulations_2023}
Röpke, F.~K., \& De~Marco, O. 2023, LRCA, 9, 2, \dodoi{10.1007/s41115-023-00017-x}

\bibitem[{Sana {et~al.}(2012)Sana, de~Mink, de~Koter, Langer, Evans, Gieles, Gosset, Izzard, Le~Bouquin, \& Schneider}]{sana_binary_2012}
Sana, H., de~Mink, S.~E., de~Koter, A., {et~al.} 2012, Science, 337, 444, \dodoi{10.1126/science.1223344}

\bibitem[{Sand {et~al.}(2020)Sand, Ohlmann, Schneider, Pakmor, \& Röpke}]{sand_common-envelope_2020}
Sand, C., Ohlmann, S.~T., Schneider, F. R.~N., Pakmor, R., \& Röpke, F.~K. 2020, \aap, 644, A60, \dodoi{10.1051/0004-6361/202038992}

\bibitem[{Smith(2011)}]{smith_explosions_2011}
Smith, N. 2011, \mnras, 415, 2020, \dodoi{10.1111/j.1365-2966.2011.18607.x}

\bibitem[{Smith(2014)}]{smith_mass_2014}
---. 2014, \araa, 52, 487, \dodoi{10.1146/annurev-astro-081913-040025}

\bibitem[{{Soker} {et~al.}(2018){Soker}, {Grichener}, \& {Sabach}}]{soker.2018}
{Soker}, N., {Grichener}, A., \& {Sabach}, E. 2018, \apjl, 863, L14, \dodoi{10.3847/2041-8213/aad736}

\bibitem[{Soker \& Tylenda(2006)}]{soker_violent_2006}
Soker, N., \& Tylenda, R. 2006, \mnras, 373, 733, \dodoi{10.1111/j.1365-2966.2006.11056.x}

\bibitem[{Staff {et~al.}(2016)Staff, De~Marco, Macdonald, Galaviz, Passy, Iaconi, \& Low}]{staff_hydrodynamic_2016}
Staff, J.~E., De~Marco, O., Macdonald, D., {et~al.} 2016, \mnras, 455, 3511, \dodoi{10.1093/mnras/stv2548}

\bibitem[{Tauris \& Dewi(2001)}]{tauris_research_2001}
Tauris, T.~M., \& Dewi, J. D.~M. 2001, \aap, 369, 170, \dodoi{10.1051/0004-6361:20010099}

\bibitem[{Tauris {et~al.}(2017)Tauris, Kramer, Freire, Wex, Janka, Langer, Podsiadlowski, Bozzo, Chaty, Kruckow, van~den Heuvel, Antoniadis, Breton, \& Champion}]{tauris_formation_2017}
Tauris, T.~M., Kramer, M., Freire, P. C.~C., {et~al.} 2017, \apj, 846, 170, \dodoi{10.3847/1538-4357/aa7e89}

\bibitem[{Thorne \& Zytkow(1975)}]{thorne_red_1975}
Thorne, K.~S., \& Zytkow, A.~N. 1975, \apj, 199, L19, \dodoi{10.1086/181839}

\bibitem[{Thorne \& Zytkow(1977)}]{thorne_stars_1977}
---. 1977, \apj, 212, 832, \dodoi{10.1086/155109}

\bibitem[{Turk {et~al.}(2011)Turk, Smith, Oishi, Skory, Skillman, Abel, \& Norman}]{turk_yt_2011}
Turk, M.~J., Smith, B.~D., Oishi, J.~S., {et~al.} 2011, \apjs, 192, 9, \dodoi{10.1088/0067-0049/192/1/9}

\bibitem[{Tutukov \& Yungelson(1993)}]{tutukov_merger_1993}
Tutukov, A.~V., \& Yungelson, L.~R. 1993, \mnras, 260, 675, \dodoi{10.1093/mnras/260.3.675}

\bibitem[{{van den Heuvel} {et~al.}(2017){van den Heuvel}, {Portegies Zwart}, \& {de Mink}}]{vandenheuvel.2017}
{van den Heuvel}, E.~P.~J., {Portegies Zwart}, S.~F., \& {de Mink}, S.~E. 2017, \mnras, 471, 4256, \dodoi{10.1093/mnras/stx1430}

\bibitem[{{van Son} {et~al.}(2022){van Son}, {de Mink}, {Callister}, {Justham}, {Renzo}, {Wagg}, {Broekgaarden}, {Kummer}, {Pakmor}, \& {Mandel}}]{vanson.2022}
{van Son}, L.~A.~C., {de Mink}, S.~E., {Callister}, T., {et~al.} 2022, \apj, 931, 17, \dodoi{10.3847/1538-4357/ac64a3}

\bibitem[{Vigna-Gómez {et~al.}(2022)Vigna-Gómez, Wassink, Klencki, Istrate, Nelemans, \& Mandel}]{vigna-gomez_stellar_2022}
Vigna-Gómez, A., Wassink, M., Klencki, J., {et~al.} 2022, \mnras, 511, 2326, \dodoi{10.1093/mnras/stac237}

\bibitem[{Vigna-Gómez {et~al.}(2018)Vigna-Gómez, Neijssel, Stevenson, Barrett, Belczynski, Justham, de~Mink, Müller, Podsiadlowski, Renzo, Szécsi, \& Mandel}]{vigna-gomez_formation_2018}
Vigna-Gómez, A., Neijssel, C.~J., Stevenson, S., {et~al.} 2018, \mnras, 481, 4009, \dodoi{10.1093/mnras/sty2463}

\bibitem[{Vigna-Gómez {et~al.}(2020)Vigna-Gómez, MacLeod, Neijssel, Broekgaarden, Justham, Howitt, de~Mink, Vinciguerra, \& Mandel}]{vigna-gomez_common_2020}
Vigna-Gómez, A., MacLeod, M., Neijssel, C.~J., {et~al.} 2020, \pasa, 37, e038, \dodoi{10.1017/pasa.2020.31}

\bibitem[{Wei {et~al.}(2023)Wei, Schneider, Podsiadlowski, Laplace, Roepke, \& Vetter}]{wei_evolution_2023}
Wei, D., Schneider, F. R.~N., Podsiadlowski, P., {et~al.} 2023, arXiv e-prints, \dodoi{10.48550/arXiv.2311.07278}

\bibitem[{{Zahn}(1977)}]{zahn.1977}
{Zahn}, J.~P. 1977, \aap, 57, 383

\end{thebibliography}
\bibliographystyle{aasjournal}

\end{document}